\documentclass[useAMS,usenatbib]{mn2e}

\usepackage{graphicx}
\usepackage{color}

\usepackage{amssymb,amsmath}

\title[Searching for solar-like oscillations in the $\delta$~Scuti star $\rho$ Puppis]{Searching for solar-like oscillations in the $\delta$~Scuti star $\rho$~Puppis\thanks{Based on data obtained at the Anglo-Australian Observatory, the European Southern Observatory, McDonald Observatory of the University of Texas and the Canadian-French Hawaii Telescope}}

\author[V. Antoci et al.]{V. Antoci$^{1,2,3}$\thanks{E-mail: antoci@phys.au.dk}, G. Handler$^{4}$, F. Grundahl$^1$, F. Carrier$^5$, E. J. Brugamyer$^6$, \and P. Robertson$^6$, H. Kjeldsen$^1$, Y. Kok$^7$, M. Ireland$^8$, J. M. Matthews$^{3}$\\
$^{1}$Stellar Astrophysics Centre, Department of Physics and Astronomy, Aarhus University, Ny Munkegade 120,\\ 
DK-8000 Aarhus C, Denmark\\
$^{2}$Institute for Astronomy, University of Vienna, T\"urkenschanzstr. 17, A-1180 Vienna, Austria\\
$^{3}$University of British Columbia, Department of Physics and Astronomy, 6224 Agricultural Rd., Vancouver, B.C., V6T 1Z1, Canada\\
$^{4}$Copernicus Astronomical Center, Bartycka 18, 00-716 Warsaw, Poland\\
$^{5}$ Katholieke Universiteit Leuven, Instituut voor Sterrenkunde, Leuven, Belgium\\
$^{6}$ Department of Astronomy, University of Texas, Austin, TX 78712, USA\\
$^{7}$ Sydney Institute for Astronomy (SIfA), School of Physics, University of Sydney, NSW 2006, Australia\\
$^{8}$ Department of Physics and Astronomy, Macquarie University, NSW 2109, Australia}

\begin{document}

\date{}

\pagerange{\pageref{firstpage}--\pageref{lastpage}} \pubyear{2002}

\maketitle

\label{firstpage}

\begin{abstract}

Despite the shallow convective envelopes of $\delta$ Scuti pulsators, solar-like oscillations 
are theoretically predicted to be excited in those stars as well. To search for such 
stochastic oscillations we organised a spectroscopic multi-site campaign for the bright, 
metal-rich $\delta$ Sct star $\rho$ Puppis. We obtained a total of 2763 high-resolution spectra using four telescopes. We discuss the reduction and analysis with the iodine cell technique, developed for searching for low-amplitude radial velocity variations, in the presence of high-amplitude variability.
Furthermore, we have determined the angular diameter of $\rho$ Puppis to be $1.68\pm0.03$ 
mas, translating into a radius of 3.52 $\pm$ 0.07~R$_{\odot}$. Using this value, the frequency of maximum power of possible solar-like oscillations, is expected at $\sim 43\pm2$~cd$^{-1}$ ($498\pm23~\mu$Hz). The dominant $\delta$ Scuti-type pulsation mode of $\rho$ Puppis is
known to be the radial fundamental mode which allows us to determine the mean density of the 
star, and therefore an expected large frequency separation of 2.73~cd$^{-1}$ (31.6 $\mu$Hz).
We conclude that 1) the radial velocity amplitudes of the $\delta$ Scuti pulsations are different for different spectral lines; 2) we can exclude solar-like oscillations to be present in $\rho$ Puppis with an amplitude per radial mode larger than 0.5 ms$^{-1}$.
\end{abstract}

\begin{keywords}

asteroseismology -- techniques: spectroscopic -- stars: chemically peculiar -- stars: individual: $\rho$ Puppis -- stars: oscillations -- stars: variables: $\delta$ Scuti.

\end{keywords}

\section{Introduction}

The $\delta$ Scuti pulsators occupy a region in the Hertzsprung-Russell Diagram (HRD) where 
many physical processes occur. For instance, it is in the classical instability strip where 
the transition between deep and efficient to shallow convective envelopes takes place. The so 
called granulation boundary (Gray \& Nagel 1989) was found from bisector measurements to 
cross following spectral types: F0V, F2.5IV, F5III, F8II, G1Ib, which intersects the $\delta$ 
Sct instability strip. Because convection has an impact not only on pulsational stability but 
also on stellar evolution, activity, stellar atmospheres, transport of angular 
momentum, etc., it is of great interest to understand how this transition takes place.

$\delta$ Sct stars are pulsating stars with masses between 1.5 and 2.5 M$_{\odot}$,  
effective temperatures ranging from 6700 to 8000 K and are situated on the pre main sequence 
(PMS), main sequence (MS) and on the immediate post MS. The periods of pulsation range from 
15 minutes to 8 hours and are the result of low radial order pressure and gravity modes, 
driven by the $\kappa$ mechanism acting like a heat engine, converting thermal into mechanical energy (e.g. Eddington 1919, Cox 1963). 
While a star is contracting, pressure and temperature increase. As a consequence the opacity ($\kappa$) will increase in certain layers
which can block the radiative flux from below. When the star returns to its state of equilibrium (due to the pressure and temperature increase) the additional stored thermal 
energy will be transformed into mechanical energy and the stellar layers will shoot past the equilibrium. While the star is 
expanding the temperature and pressure decrease and so does the opacity, the latter is because the ionized gas is more 
transparent. The star will stop expanding and contracts, the ions will recombine and $\kappa$ increases again and the cycle repeats. 
The specified layers where the thermal energy is stored, are the zones connected to (partial) ionization of abundant elements 
which can take place only at specific temperatures. The zone where neutral hydrogen (H) and helium (He) are 
ionized has about 14 000 K and is close to the surface. The second ionization zone of He II is at $\sim$50 000 K and drives the pulsation in $\delta$ Sct stars. 
The location in the HRD is confined within a region defined by the so called red (cool) 
and blue (hot) edges, respectively. Stars with effective temperatures higher than the blue 
edge do not pulsate because the He II opacity bump is at a depth where the density is too low 
to excite pulsation (Pamyatnykh 2000). Near the cool border of the instability strip, the 
oscillations are damped due to the increasing depth of the convective outer layers. This 
border of the instability strip was theoretically calculated by Dupret et al. (2005), using 
time dependent convection, which requires a non-adiabatic treatment that considers the 
interaction between pulsation and convection as described by Grigahc\'ene et al. (2005). 
Interestingly, the instability domain of the $\gamma$ Doradus ($\gamma$ Dor) stars, pulsating 
in high radial order g modes, overlaps that of the $\delta$ Sct stars, suggesting that both 
types of pulsation should be excited simultaneously in stars in the overlap region (Handler 
1999). Satellite missions like MOST, CoRoT and {\it Kepler} have established that the $\delta$ Sct 
and $\gamma$ Dor hybrids are a common phenomenon, rather than an exception (Grigahc\'ene et 
al. 2010, Uytterhoeven et al. 2011).

In earlier days, models of $\delta$ Sct stars predicted many more frequencies to be
excited than were observed. Extensive ground-based observing campaigns showed that lowering 
the detection level is imperative to finding additional independent modes (e.g., Breger et al. 
2005). Nowadays, thanks to asteroseismic space missions, in some cases ``too many'' 
frequencies are observed. For instance, Poretti et al. (2009) needed around 1000 frequencies 
to fit the CoRoT light curve of the $\delta$ Sct star HD 50844. Mode identification based on 
spectroscopic data suggested that modes with high spherical degree ($\ell$=14) are excited. 
Several radial orders of high spherical degree modes split by rotation would be required to 
explain the large number of observed frequencies. A second CoRoT star (HD 174936, Garc\'{i}a 
Hern\'{a}ndez et al. 2009) was found to have approximately 500 significant frequencies 
attributed to pulsation.

Kallinger \& Matthews (2010) suggested that the major part of the observed frequencies 
originates from granulation noise and is not due to pulsation. Their interpretation assumes 
the presence of non-white granulation noise, as observed in sun-like stars and red giants. 
The granulation time scales found in the two CoROT stars are consistent with scaling 
relations from cooler stars. The granulation hypothesis is strongly supported by earlier 
spectroscopic measurements, where line profile variations were found to be caused by 
convective motions in the atmosphere of A and Am stars (Landstreet et al. 2009). In general, 
A type stars have rather high microturbulent velocities\footnote{This parameter mirrors the 
velocity fields in the stellar atmosphere, hence can be related to convection.} which is 
further evidence for granulation (Montalban et al. 2007). Balona (2011) confirmed that in the 
{\it Kepler} A type stars (whether variable or not) there is a correlation between the noise level 
and the effective temperature as expected for the signature of convection. However, it is plausible that 
we observe a combination of both high order $\ell$ modes 
and granulation noise. The data quality from space is indeed high enough to also observe 
higher $\ell$ modes, as already shown by Balona \& Dziembowski (1999).
These authors calculated that the amplitudes should drop below 50 
$\mu$mag for modes with $\ell\geq6$ due to geometrical effects, assuming the same intrinsic 
amplitude of $\Delta T_{\rm eff}/T_{\rm eff}\approx0.01$ for all modes. This is a value which 
can be detected from space.

Even given the present possibilities to detect large numbers of pulsational mode frequencies, 
unique seismic modelling of $\delta$ Sct stars is still hampered by many obstacles. The major 
difficulty concerns the identification of the pulsation modes. It is not yet known why some 
of the modes observed in $\delta$ Sct stars are excited to observable amplitudes while others 
are not. To clearly determine the geometry of a mode, long observing campaigns, multicolour 
photometry and/or high-resolution time-resolved spectroscopy are needed.

Houdek et al. (1999) and Samadi et al. (2002) predicted that the convection in outer layers 
of $\delta$ Sct stars is still efficient enough to excite solar-like oscillations. Using 
time-dependent, nonlocal, mixing-length convective treatment, they calculated the intrinsic 
stability of modes excited by the $\kappa$ mechanism. Only when modes are intrinsically 
stable acoustic noise generated by turbulent convection can excite the natural eigenmodes in 
a given frequency range. Contrary to the modes excited by the $\kappa$ mechanism, the 
amplitudes of stochastically excited oscillations can be approximated by theoretical and 
empirical scaling relations as they are determined by the interaction between excitation and 
damping. Convection excites {\it all} the modes in a certain frequency regime, again in 
absolute contrast to the $\kappa$ mechanism, allowing mode identification from pattern 
recognition. This is possible because consecutive high radial order p modes are approximately 
equally spaced in frequency, which gives rise to a comb-like structure in periodograms. The 
large frequency separation $\Delta\nu$ is the mean separation between high radial order modes 
of the same degree $\ell$ and measures the mean density of the star. The stochastic 
oscillation spectrum is described by a broad envelope with a maximum at the frequency of 
maximum power $\nu_{max}$, which is related to the acoustic cut-off frequency. A very tight 
scaling relation relation between these two parameters, $\Delta\nu$ and $\nu_{max}$ allows to 
derive basic stellar parameters, such as radius and mass (cf. Kjeldsen \& Bedding, 1995, 
Stello et al. 2009, Huber et al. 2011).

All these tight relations, even if not perfect in some cases, allow to perform 
seismology of solar-like stars with high accuracy. The derived mass distribution of the 
{\it Kepler} population of Sun-like stars may have a large impact on the models predicting the 
population of our Galaxy (Chaplin et al. 2011). Can the same be done for $\delta$ Sct stars? 
Are there relations we can use to study these stars as a group? If so, Uytterhoeven et al. 
(2011) and Balona (2011) who statistically analyzed all the A and F stars in the {\it Kepler} FOV, 
at least did not find any.\par
Recently, Antoci et al. (2011) reported the first evidence for solar-like oscillations in the 
$\delta$ Sct star HD 187547 observed by the {\it Kepler} spacecraft, but in no other member 
of the same type. New observations, however suggest that the situation is more complicated than 
anticipated. It is true that all sun-like stars observed with sufficient accuracy show 
solar-like oscillations but should all the $\delta$ Scuti stars behave alike?\par
Sun-like stars are, compared to more massive stars like the $\delta$ Sct stars, more 
homogeneous when it comes to rotation for example. The stars located in the $\delta$ Sct instability strip 
have rotational velocities from very slow (44 Tau has a projected rotational velocity, v sin 
i = 2 $\pm$ 1 kms$^{-1}$, Zima et al. 2009) to almost break-up velocity (Altair, e.g. Monnier 
et al. 2007). In the case of Altair the authors find a difference in temperature between the pole and the 
equator of more than 1500K (!). Fast rotation, for example, (also common in $\delta$ Sct stars, e.g. Solano \& Fernley (1997))
combined with differential rotation is suggested to influence the depth of the convective 
outer layers such that solar-like oscillations might be excited in stars in which otherwise 
the convection would not be effective enough (MacGregor et al. 2007). On the other hand, 
theoretical modelling of the stars FG Vir (Daszy\'{n}ska-Daszkiewicz et al. 2005) and 44 Tau 
(Lenz et al. 2010) suggests insignificant convection in their envelopes. \par
This  emphasizes the {\it inhomogeneity} of $\delta$ Sct stars
which may cast some doubts on whether {\it all} stars with a given temperature 
and luminosity should show solar-like oscillations. Carrier et al. (2007) 
have searched for solar-like oscillations in the spectroscopic binary Am star HD 209625 
(which is not classified as a $\delta$ Sct star), but did not detect any. \par
The main motivation for the present work were the theoretical predictions that solar-like 
oscillations should be present in $\delta$ Sct stars (Houdek et al. 1999, Samadi et al. 
2002). If detected, that would allow us to perform mode identification for the stochastically 
excited oscillations from pattern recognition and therefore to also identify the modes of 
pulsation excited by the $\kappa$ mechanism. As a result we would be able to model the deep 
interior of the star by using the $\delta$ Sct oscillations and constrain the overshooting 
parameter $\alpha_{OV}$, which is the primary factor to influence the main sequence lifetimes 
of stars with convective cores. The solar-like oscillations on the other hand 
are most sensitive to the outer (convective) envelope, which so far is poorly understood in stars 
with masses typical of $\delta$ Sct stars. To search for solar-like oscillations in 
$\delta$ Sct stars, we chose in this case to use radial velocity (RV) measurements. The reason is 
that the noise due to granulation is lower in RV measurements than in photometry, in the solar case between 10 and 30 times lower, depending on the frequency 
(e.g., see Fig. 4.1 by Aerts et al. 2010).

\subsection[]{$\rho$ Puppis}

$\rho$ Puppis (HD 67523, HR 3185, spectral type {\it k}F3V{\it h}F5{\it m}F5; Gray \& Garrison 1989) is a bright star with V = 2.81~mag (RA$_{2000}=08~07~32.6489$, 
DEC$_{2000}=-24~18~15.567$) and was target of many studies as it is eponymous for the 
metal-rich $\delta$ Sct stars and is the brightest of its class.

It was first discovered to show light variability by Eggen 
(1956) and was classified as a $\delta$ Sct star. Fracassini et al. (1983) found strong 
evidence for emission in the Mg II H + K lines ($\lambda \lambda$ 2802.7 and 2795.5) present 
during the whole pulsation cycle which is a signature of chromospheric activity. Yang et al. 
(1987) determined different amplitudes of pulsation for the dominant mode in different 
elements in the spectra of $\rho$ Puppis which they interpreted to be most probably due to 
beating effects with other pulsation modes. The same authors reported a possible van Hoof 
effect\footnote{The pulsation amplitudes and/or phases measured for different elements can 
significantly differ, caused either by shock waves or by running waves (see e.g. Mathias \& 
Gillet 1993)}. Mathias et al. (1997) were the first who detected two additional frequencies in new radial velocity measurements, increasing the number of pulsation modes to three ($\nu_0$=7.098168~cd$^{-1}$, $\nu_1$= 7.815~cd$^{-1}$ and $\nu_2$=6.314~cd$^{-1}$ respectively). Using the 
moment method (e.g. Aerts 1996) the authors identified the dominant mode as a radial mode and the second one as 
an axisymmetric one. Mathias et al. (1997) also found evidence for a standing wave effect 
between Fe I $\lambda \lambda$ 6546 and the H$\alpha$ line but did not observe any signature 
of activity in the Ca II K line as reported earlier by Dravins et al. (1977). Mathias et al. 
(1999) re-observed $\rho$ Puppis to test the hypothesis of the transient feature but again 
did not find any evidence.

Antoci et al. (2013) discuss literature determinations of the position of $\rho$ Puppis in the HR Diagram. Following these authors, we adopt $T_{\rm eff}=6900\pm150$~K and $\log g=3.8\pm0.2$. There is an accurate HIPPARCOS parallax of $51.33 \pm 0.15$ mas (van Leeuwen 2007) for the star, which in combination with the effective temperature implies an evolutionary status somewhat off the main sequence and a mass of about 1.8~M$_{\sun}$.


\section[]{Observations}


For the present work we organized a spectroscopic multi-site campaign (summarized in Table 1) resulting in 2763 high signal to noise high-resolution spectra.

\begin{table*}

\caption{Spectroscopic multisite campaign for $\rho$ Puppis. }


\begin{tabular}[h!]{llllccc}\hline

Telescope & Instrument & WL calibration& RV extraction & year-month & Nr. of  & Nr. of  \\
            &            &        &       &     & nights  & measurements
 \\\hline
1.2-m Swiss telescope & CORALIE          & ThAr &CCF & 2008-01 & 4   & 671    \\
AAT      & UCLES            & $I_{2}$ cell & iSONG   & 2008-01 & 5   & 1165   \\
2.7-m telescope@McDonald &  Robert G. Tull    & $I_{2}$ cell & iSONG   & 2010-01 & 4   & 424    \\
& Coud\'e Spectrograph  & & & & & \\
CFHT     & ESPaDOnS         & ThAr  &bisectors & 2010-03 & 3   & 503    \\
& &+ telluric lines & & & & \\
\hline
 \end{tabular}
\label{data}



\end{table*}\par

During January 2008 we obtained data with the CORALIE ($R = 50000$, resolution element of 6~kms$^{-1}$) spectrograph mounted 
at the 1.2-m Swiss telescope at ESO, observer being FC. These data were 
reduced using the pipeline available for the spectrograph and the RV measurements were 
obtained by cross-correlating the observed spectra with a mask represented by a synthetic 
spectrum. Simultaneous observations to minimise aliasing due to daylight gaps were 
undertaken at the 3.9-m AAT (Anglo Australian Telescope) with the UCLES spectrograph and through an iodine cell, 
spread over 5 nights (observers VA and GH). The exposure times varied between 45 and 120 s to 
account for seeing variations during the night. With a resolution R=48000 and a resolution element of 6.25~kms$^{-1}$, the echelle 
spectra consist of 47 orders, of which 30 cover the wavelength region of the iodine cell. 
Measurements at McDonald observatory were acquired by EJB and PR from January 27 to February 1, 2010 
with the Coud\'e Spectrograph at the 2.7-m telescope, at $R=60000$ (resolution element of 5~kms$^{-1}$), also using an iodine 
cell.  The AAT and McDonald data were reduced applying standard procedures using IRAF, i.e., bias correction, flat-fielding, scattered light subtraction, order fitting and tracing and rough normalisation. Observations with the 
ESPaDOnS spectrograph at the 3.6-m CFHT (Canada-France-Hawaii Telescope) were gathered in service mode from March $4-7$, 2010 and reduced 
with the {\it Libre Esprit} pipeline (Donati et al. 1997). The spectral resolution in the star-only mode was 81000 (resolution element of 3.7~kms$^{-1}$), and 
wavelength calibration was done using the ThAr frames as well as telluric lines.

\subsection[]{Radial velocity measurements using the iodine cell technique}

The iodine cell method, developed by Marcy \& Butler (1993) and Butler et al. (1996), was initially used to detect 
Doppler shifts caused by planets orbiting their host stars. It is based on the idea that 
the stellar light and the reference frame should follow the same path through the 
telescope and the spectrograph, minimizing temporal and spatial error sources on the 
radial velocity measurements. The cell is mounted in front of the spectrograph. The 
reason for choosing $I_{2}$ is that it shows a forest of absorption lines between 5000 
and 6000\,\AA, a region where stars with a spectral type later than A have many spectral 
lines providing the necessary information for precise radial velocity measurements. The 
wavelengths of the $I_{2}$ absorption lines are very well determined by scanning the cell 
with a Fourier Transform Spectrometer (FTS).

For the search of signals with tiny amplitudes of the order of few meters per second a 
complex reduction method, involving the modelling of the instrumental 
point-spread-function (PSF) of the spectrograph is needed. The shape of the instrumental 
profile is influenced by many sources, such as seeing changes during the night, optical 
elements, dust, etc. A change of only 1\% in the PSF shape, which is normally 
approximated by a Gaussian, can introduce shifts in the radial velocity measurements of 
several ms$^{-1}$ (Butler et al. 1996).

During the whole reduction process the echelle orders are divided into chunks of several 
pixels, covering $2-2.5$\,\AA, each delivering a radial velocity measurement. The 
wavelength region covered by the $I_{2}$ spectrum is $\sim$1000\,\AA~ spread over several 
orders. Depending on the dispersion of the spectrograph and the CCD detector a number of 
200-300 of `individual' radial velocity measurements can be obtained allowing a high 
precision of the final value. To derive the radial velocity of a chunk the model is 
defined by the product of the intrinsic stellar spectrum $I_S$ and the transmission 
function of the iodine absorption cell T$_{I_2}$, which then is convolved with the 
spectrograph's PSF (usually ten to twenty times smaller than the chunk size). The result is afterwards binned to the wavelength range of the CCD. 
The stellar spectrum  observed through the  $I_{2}$ cell, $I_{obs}$, is described as follows

\begin{equation}\label{Iobs}
I_{obs} = k[T_{I_2}(\lambda)I_S(\lambda + \Delta \lambda)]*PSF
\end{equation}

where $k$ is a normalization factor proportional to the exposure level (cf. Marcy \& Butler 1993), $\Delta \lambda$ the Doppler shift and $*PSF$ 
represents the convolution with the PSF of the spectrograph (Butler et al. 1996). 
$T_{I_2}$, the transmission function of the spectrum of the iodine cell is the FTS spectrum obtained as 
described above. The intrinsic stellar spectrum $I_S$ must resemble the spectrum free of 
any instrumental effects, which is done by deconvolution of the target spectrum (not observed 
through the iodine cell) and the PSF. For the latter, a rapidly rotating OB type star is 
observed through the iodine cell, mimicking a calibration lamp. It can be assumed that 
the stellar spectral features are smeared out, and the iodine spectrum is imaged. From 
that, the PSF of the instrument can be derived. This procedure is more accurate than 
measuring a real calibration lamp, because the calibration star light follows the same 
path as that of the target star. The PSF itself is a sum of several Gaussians at fixed 
position and widths but with variable height. In this way a possible non-symmetric PSF 
can be described. For a detailed explanation on how to derive a PSF see Valenti et al. 
(1995). To preserve the detailed wavelength information provided by the $I_{2}$ cell, the 
stellar spectra are oversampled by a certain factor (at least 4 times) to which also the 
$I_{2}$ cell is resampled. Now having the deconvolved intrinsic stellar spectrum $I_S$, 
the PSF of the instrument and the iodine cell spectrum, the observations can be modelled 
as described in Equation \ref{Iobs}. For this final step the Doppler shift $\Delta 
\lambda$ becomes an additional free parameter, the one of interest which allows 
to reproduce the observations. Figure 1 by Butler et al. (1996) nicely illustrates 
the procedure described above.

The {\it Stellar Observations Network Group} (SONG, Grundahl et al. 2011) is an ongoing 
project which aims to establish a global network of eight robotic 1-m telescopes equipped 
with high-resolution spectrographs. To obtain ultra-precise radial velocity measurements 
the iodine cell method was chosen. During the planning phase of the SONG project a code 
(iSONG) to extract precise radial velocity measurements of a star observed through an 
iodine cell was developed by FG and is based on the method described above 
(Butler et al. 1996). A direct comparison (Grundahl et al. 2008) using the same data set 
of $\alpha$ Cen A observed with UVES@VLT, demonstrates that iSONG is capable of reaching 
nearly the same precision as mentioned by Butler et al. (1996). To analyse the data for the present 
work we used iSONG.

\subsubsection{Determining the instrumental profile}

To obtain the instrumental profile of UCLES, we observed the rapidly rotating early B 
type star HD~64740, whereas for the McDonald spectrograph we observed the rapidly 
rotating B type star HD~67797. The PSF of each spectral chunk was calculated by first 
approximating the instrumental profile by a single Gaussian with a fixed position but a variable height and width ($\sigma$). The latter  
values were then used as an initial guess for the multiple Gaussian. For UCLES and the McDonald spectrographs we find a mean PSF of 3.2 and 2.4 pixel widths respectively. These values are significantly lower than the size of the chunks which were chosen to be 71 pixels in both cases.

Even if the B type 
stars used to derive the instrumental PSF had very few absorption lines, some were still 
present. The chunks covering those wavelengths regions could not be used. In some other 
cases we obtained unusual-looking Gaussians as well, but with no obvious reason. To avoid 
losing these chunks ($\sim30\%$), we replaced them by the mean PSF computed from the reliable chunks.

\subsubsection{Intrinsic stellar spectrum}

In the case of iSONG the deconvolution is done by using the procedure described by Crilly 
et al. (2002). To model the intrinsic stellar spectrum a very high S/N spectrum of the 
target star, not taken through the iodine cell, is needed. To this end, usually many 
stellar templates without iodine lines are combined. In the case of the AAT data we only 
had few and all were from different pulsation phases. As there are equivalent width (EQW) 
variations due to the high amplitude of pulsation of the dominant $\delta$~Scuti-type 
pulsation ($\sim$ 10 kms$^{-1}$ peak-to-peak amplitude), it turned out that not combining 
our template spectra resulted in less scattered RV measurements. To have consistent data sets, 
the same was done for the McDonald measurements. Finally we chose the spectra with a 
phase closest to the {\it minimum change} in EQW. Butler et al. (1993) observed a Cepheid which 
has very large changes in EQW during a pulsation cycle using the same technique. They 
chose to use more templates, each observed in a specific phase of pulsation. In the case 
of $\rho$ Puppis however, the average difference in RV when using templates observed at different phases of pulsations is minimal, of the order of $1-3$ ms$^{-1}$ per 
spectrum. Both, the AAT and McDonald template spectra have a S/N of 200 and higher.

\subsubsection{Fitting the observations}

To derive the RV we used the instrumental profile, the intrinsic stellar spectrum and the 
information provided by the iodine absorption lines as described above. As an initial 
guess we used the RV obtained by cross-correlating one order without $I_2$ absorption 
lines of each of the science spectra with the same wavelength region of the template.

\begin{figure}
\begin{center}
\includegraphics[width=8cm]{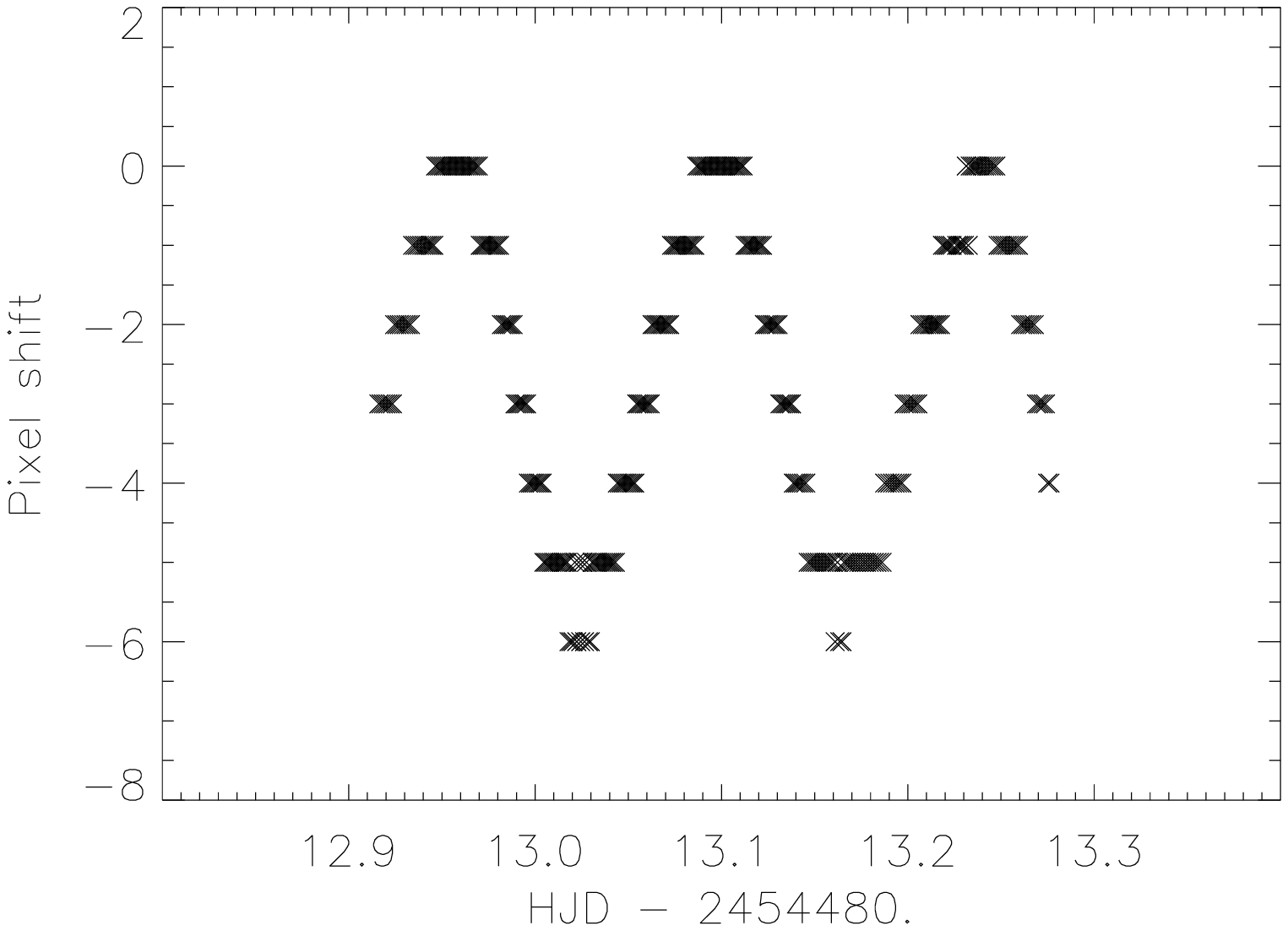}
\includegraphics[width=8cm]{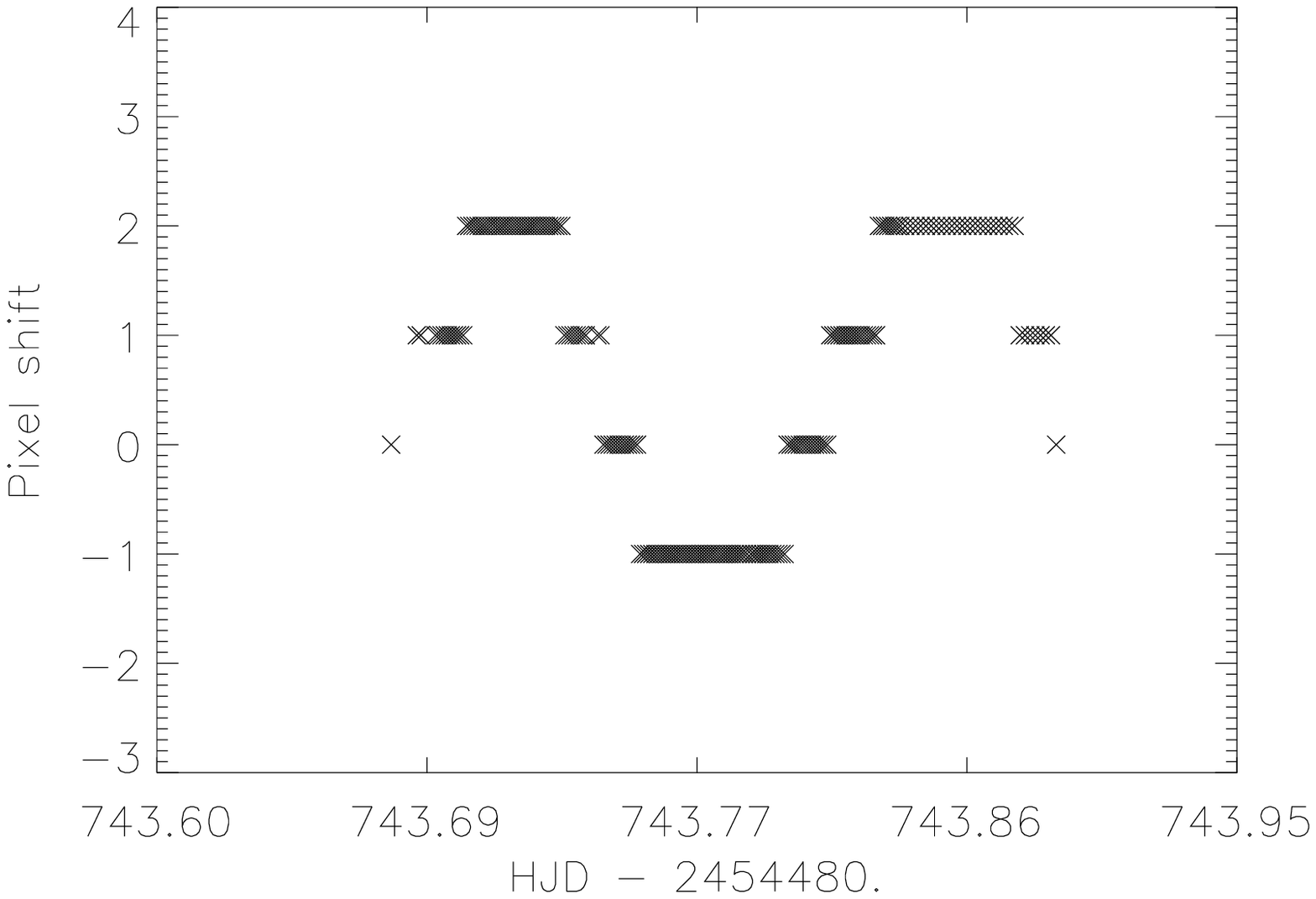}
\end{center}
\caption{Upper panel: Pixel shift relative to the template (AAT  data). Lower panel: Pixel shift relative to the template (McDonald data). }
\label{pixelshift}
\end{figure}

Because the dominant mode is radial, during a pulsation cycle the lines are periodically 
blueshifted and redshifted.  This means that a given spectral line is captured on a different pixel.  
If the chunk would be fixed at the same position on the CCD the spectral lines would move 
in and out and the actual RV measurement would always reflect different parts 
of the spectrum. In the case of $\rho$ Puppis, we first determined the pixel shift induced by the pulsation for every measurement using the cross-correlated RV. This 
information was then used to shift each chunk such that it measures the same wavelength region regardless of the pulsation phase. Experiments have shown that the scatter decreased when deriving the RV using this 'sliding window' technique. In Fig. 
\ref{pixelshift} we show the shift in pixels relative to the first spectrum for the AAT 
and McDonald data respectively.

Another parameter that influences the results is the oversampling factor. It is usually 
set to 4 or higher (Butler et al. 1996) and determines by how much the stellar spectra 
are oversampled with respect to the spectral resolution, to make use of the wavelength 
information provided by the $I_2$ cell. The latter is also resampled to the same scale. 
Increasing this factor also increases the computing time significantly, and can last up 
to hours for only one science frame. In the case of AAT data, for example, with a total 
number of 1165 spectra the computations would take several weeks for an oversampling factor of 16 or higher! Tests with different 
oversampling factors, 4, 8 and 16, have shown that 8 is the best in keeping the balance 
between a satisfactory quality and acceptable duration of computing time, which was 
also chosen for all further data analyses.

\begin{figure}

\begin{center}

\includegraphics[width=8cm]{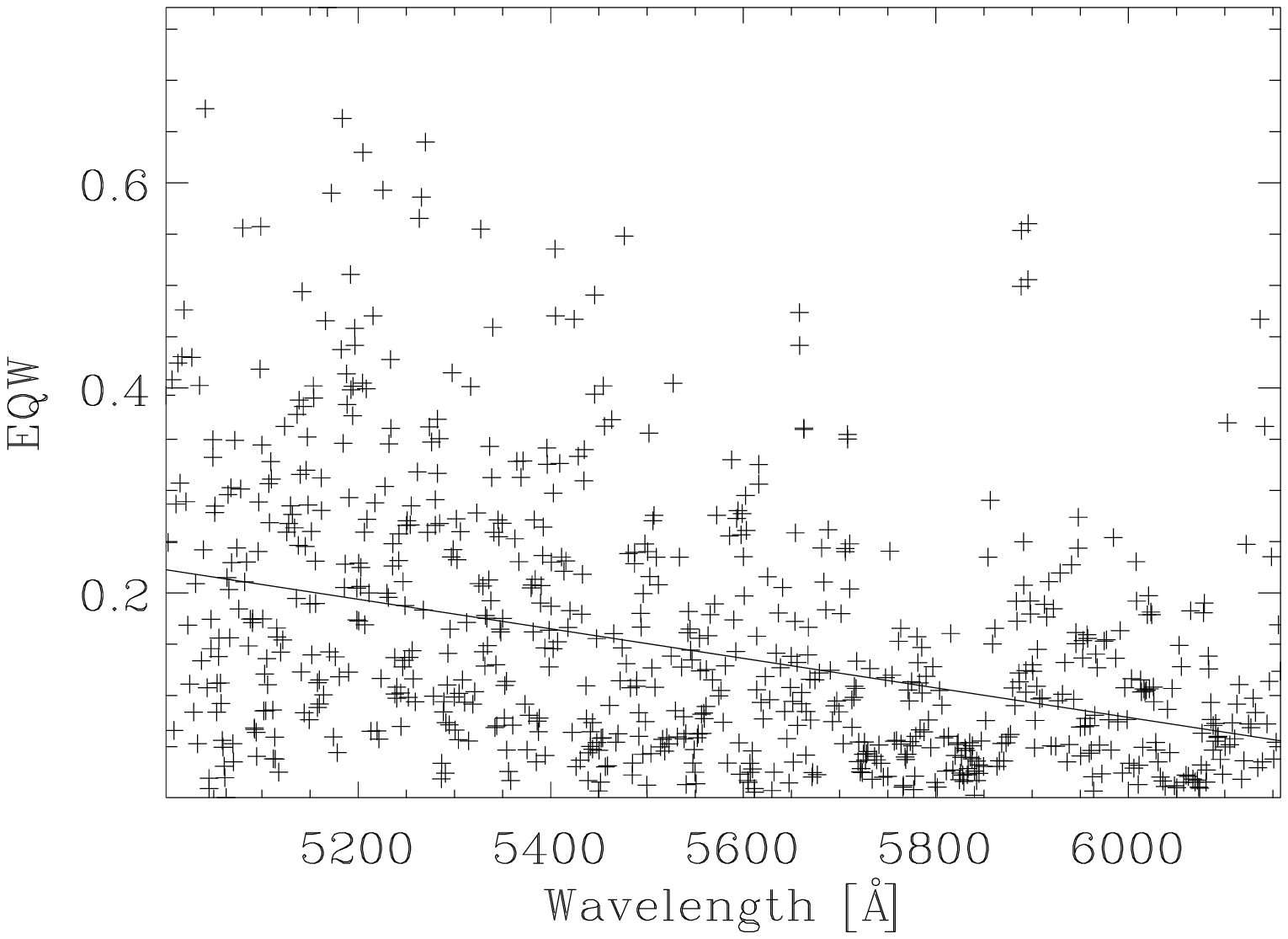}

\includegraphics[width=8cm]{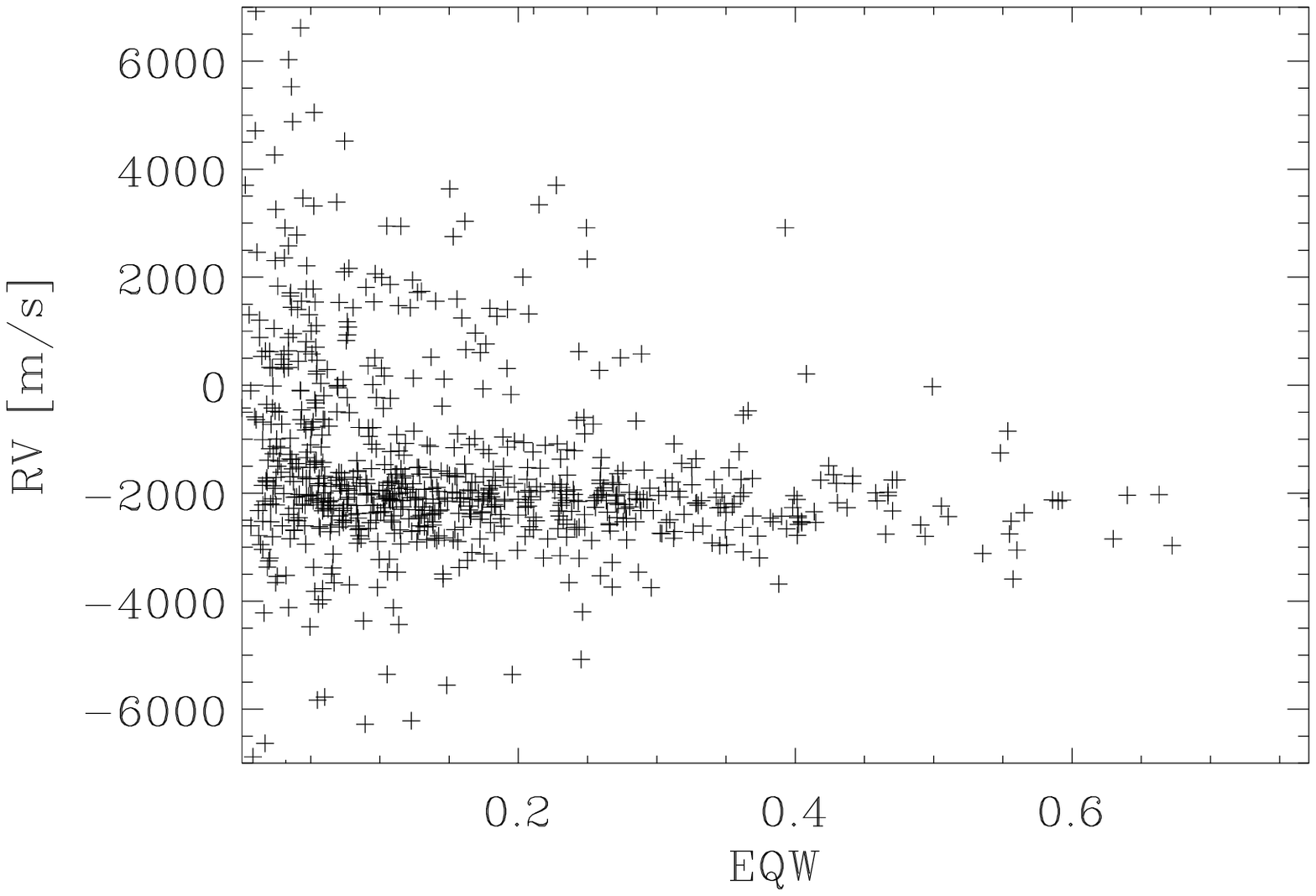}

\caption[]{Upper pannel: EQW in a chunk as a function of wavelength of one measurement (AAT  data). Lower panel: RV as a function of EQW in a chunk.  }

\label{zzzz217WLRV}

\end{center}

\end{figure}

Once the modelling was finished, each spectrum delivered 21 orders $\times$ 36 chunks = 756 
``independent'' RV measurements in the case of the AAT data, whereas each McDonald science 
spectrum consisted of 13 orders $\times$ 22 chunks = 286 RVs. Excluding the outliers and 
combining the remaining values to their mean should then result in a very precise RV 
measurement. Unfortunately, identifying the outliers in an objective way is not 
straightforward. More importantly, we also tried to separate the factors which mostly 
influence the measurement precision: RV information originates from stellar absorption lines that 
are not uniformly distributed across the spectrum. The wavelength region that can be used 
is restricted to approximately $5000-6200$\,\AA~ when using the $I_2$ technique. There are 
several chunks which do not show any stellar lines and therefore are not expected to 
carry any useful RV information. Also the density of the $I_2$ lines is getting lower when 
approaching the wavelength limits given above. The upper panel of Fig.~\ref{zzzz217WLRV} 
shows the EQW\footnote{The EQW for a given chunk was measured by integrating the area 
covered by absorption lines.} of a given spectrum as a function of wavelength as well as 
a linear trend, clearly illustrating that the number of lines decreases with increasing 
wavelength. This is consistent with what is expected for A and F type stars. Nevertheless 
it should be kept in mind that at higher wavelength also telluric lines start to appear 
in the spectrum, in other words a high EQW does not necessarily mean a stellar line. This 
is illustrated in the lower panel of Fig.~\ref{zzzz217WLRV}, where the RV measurements 
of each chunk of a given spectrum are plotted as a function of the EQW of the chunk. There is 
a clear trend showing that chunks having a higher EQW cluster around a RV value of 2000 
ms$^{-1}$, but we can also see outliers.

The goodness of the fit is expressed by the rms value and represents the residuals between 
the observations and the model in percent. In principle, the lower the value the better 
the model fits the observations. We used the measurements with a cutoff at a sigma fit value of $\sim0.25$, only few 
outliers can be found above. Chunks with an EQW larger than $\sim0.2$ also tend to have a 
larger sigma. As already mentioned a large EQW does not necessarily mean a stellar line, 
but can be attributed to telluric lines or to a mixture of both.\par

The errors per measurement (i.e. per spectrum) were estimated by calculating the rms 
scatter. To obtain the lowest error and to find out which parameter has the largest impact 
on the quality of the measurements, experiments varying three different parameters 
restricting the number of chunks used, were undertaken. The errors per measurement were 
then calculated for each spectrum using the RV information from the selected chunks 
depending on EQW, wavelength range, rms value. To decide which parameter constellation delivered 
the best data set, a ``lowest mean error'' was introduced, by averaging the errors per 
measurement for the entire data set (for AAT 1165 errors). The lowest mean error (14.5 
ms$^{-1}$) was found when all the chunks with wavelengths between 5000 and 5900\,\AA, EQW $\ge$ 0.2 
and with sigma fit value $\le$ 0.1 were used. The minimum EQW seems to have the largest 
impact on the errors, which is not surprising because chunks with very low EQW values 
carry little or no RV information.

\begin{figure}

\begin{center}

\includegraphics[width=8cm]{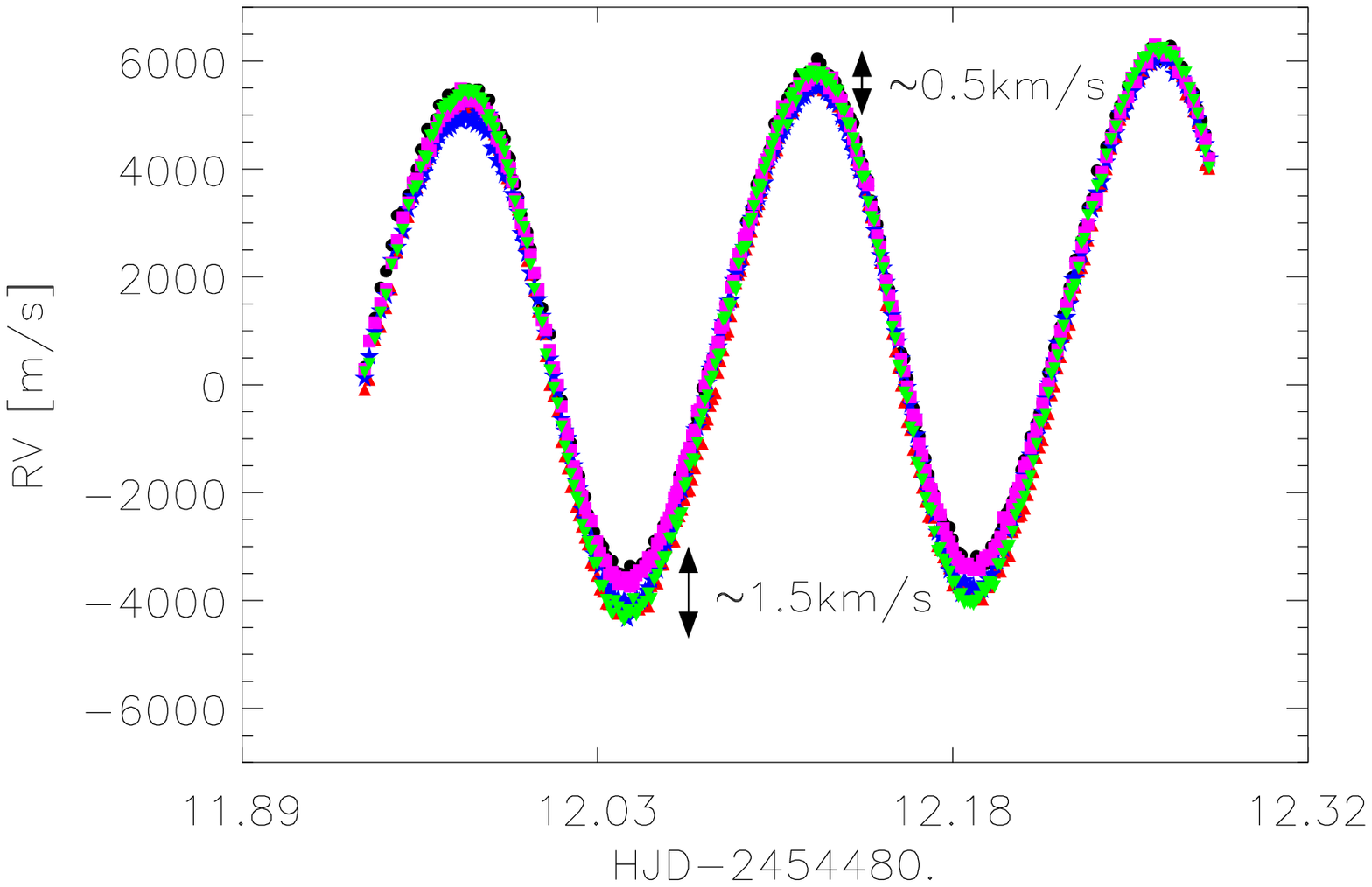}

\includegraphics[width=8cm]{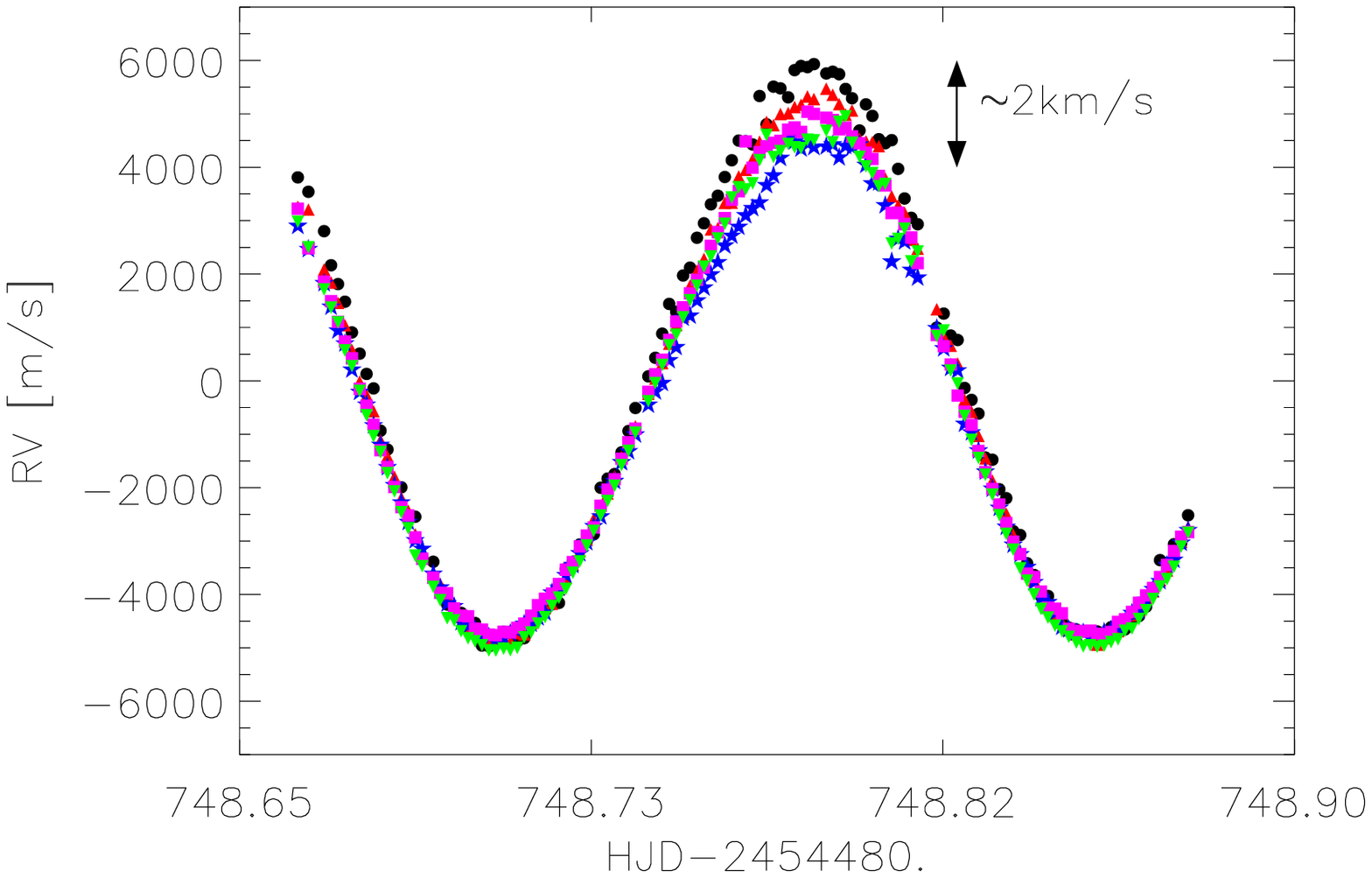}

\caption[]{Upper panel: AAT RV measurements for one night derived from different chunks. There are clear variations in amplitude. Lower panel: same plot but for the McDonald data. }

\label{rv_diffchunks}

\end{center}

\end{figure}

Even if there were systematic effects which might have been introduced by the lack of absorption lines or by strong telluric lines in some chunks or by any other effect, 
we concluded that the amplitude of the dominant mode intrinsically varies from chunk 
to chunk, i.e. from line to line. Amplitude variability (from chunk to chunk) is also present in the McDonald data and the ESPaDOnS measurements, the latter data set was not observed through an iodine cell and therefore was not reduced using iSONG. Further support for amplitude variability comes also from literature: Yang et al. (1987) and Mathias et al. (1997).
An example of several RV curves derived from different chunks for the AAT as well as for the McDonald data can be found 
in Fig. \ref{rv_diffchunks}. The amplitudes differ up to 2000 ms$^{-1}$ from peak to peak 
amplitude which translates to approximately 20\% of the pulsation amplitude. 
Such large differences in amplitude preclude a possible 
detection of the predicted solar-like oscillations with expected amplitudes between 1-2 ms$^{-1}$ if averaging over all chunks, even after careful selection. \par

In the end it turned out that an additional visual inspection of all the  ``best'' RV curves, selected using the method described above,
(AAT and McDonald data) was the best method to determine which chunks to keep for further analyses. 
We 
selected the best curves based on their scatter, some illustrated in Fig. 
\ref{rv_diffchunks}, which can be done visually simply because the variability of $\rho$ Puppis is clearly visible. This is very different in the case of signals with small amplitudes, originating from, e.g., solar-like oscillations or planetary companions, where a statistical approach is the only possible method to quantify the quality of the data set.

After the barycentric correction was done, the dominant frequencies 
derived photometrically (see section 3) where prewhitened (the amplitudes and phases 
were free parameters) from each of the RV curves independently and only then the residuals 
where combined allowing to increase the precision significantly (see section on analyses for details). \par

\subsection[]{ESPaDOnS data}

The radial velocities measured with ESPaDOnS were extracted using bisectors, which are 
usually used to study line asymmetries. The latter can be caused by different mechanisms  
such as pulsation, stellar rotation, magnetic fields, mass loss, granulation, etc.  A bisector 
is constructed by connecting the blue and the red line component 
of an absorption line at the same height and taking its middle point. If the line is 
symmetric a vertical straight line will be the result, whereas non-symmetric lines will 
cause different shapes depending on the velocity fields occurring in the photosphere of a 
star. This method has been extensively used to, e.g., study convective motions in cool 
stars (see e.g. Gray 2005) as well as to map the photospheres of roAp stars (e.g. Elkin et 
al. 2008). This can be done because the core of an absorption line is formed in higher 
photospheric layers relative to its wings. The shape, the span of velocity as well as the 
absolute position in wavelength of a bisector display important depth information (see, 
e.g., Gray 2009). \par
It is out of scope of the present paper to disentangle the velocity fields originating 
from pulsation from those caused by granulation. This means that we were not necessarily restricted to 
unblended lines, which are normally used when line asymmetries are studied, but could also 
use lines blended by the same element with the same contribution\footnote{ i.e. same abundance and excitation potential, which means that it can safely be assumed that the lines form in the same atmospheric layer and therefore have the same line profiles and Doppler shifts} to the line as we only want to determine the RV. This procedure might influence the amplitude of non-radial modes, but for radial modes it should be safe because the line will periodically blue and red shift with the pulsation period.
For line 
identification we used the synthetic spectra computed for the abundance analyses of $ 
\rho$ Puppis as described by Antoci, Nesvacil \& Handler (2008). The code to derive the line 
bisectors was kindly provided by Luca Fossati and was written in IDL. \par

\begin{figure}
\begin{center}
\includegraphics[width=8cm]{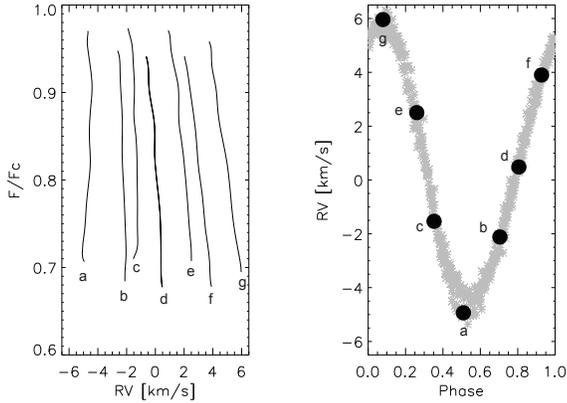}
\caption[]{Left panel: Bisectors from different phases of pulsation. The differences in amplitude at different relative fluxes can be as large as 2000 ms$^{-1}$. Right panel: Phase plot of the dominant radial mode. The black dots illustrate the phases at which the bisectors in the right-hand panel were chosen. }

\label{bis}

\end{center}
\end{figure}

After analysing several lines we found a strong RV amplitude dependence on relative fluxes.
As the data were reduced using the {\it Libre 
Esprit} pipeline available for the ESPaDOnS instrument, which has been tested extensively, 
it is very unlikely that we are dealing with artefacts. The normalisation of the 
spectrum has a very high impact on the bisector, because it can introduce or cancel 
asymmetries in the spectral line. However, our results should not be affected by this because, firstly, we do not use the part of the bisector close to the continuum (see Fig.~\ref{bis}). Secondly, to explore this problem we have tested lines which have a clear continuum to the blue and red part of the line,  e.g. the Fe I line at $\lambda \lambda$~6252~\AA, and find the same results.  Given the high quality of our data (SNR per CCD pixel at e.g. 5520~\AA~ is around 810) and comparing the observed spectrum with a synthetic spectrum for a star with the parameters of $\rho$ Puppis defining the continuum was a straightforward procedure.

In fact Mathias et al. (1997) already reported amplitude variability as function of line depth for the Ca II  $\lambda \lambda$~6122~\AA~line. 
This phenomenon can originate from a running wave character of the dominant pulsation and is sometimes also referred to as  
the Van Hoof effect, although the latter can have different interpretations (Matthias et al. 1997 and references therein). The running wave scenario reflects the propagation of the pulsation through the stellar photosphere.
Figure \ref{bis} illustrates several bisectors at different phases of the dominant mode of pulsation (7.09823 cd$^{-1}$). The differences in RV as a 
function of the line depth can be as large as 2000 ms~$^{-1}$ at maximum velocity change (indicated by the label (g) in Fig.~\ref{bis}). 
The shape of the bisector at the other extreme, on the other hand, is different and has also a different RV gradient (label (a) in Fig.~\ref{bis}). 
Because of their limited data set Mathias et al. (1997) could not rule out that non-radial modes were the cause for  this phenomenon. 
Our extensive photometric campaign, however, allowed us to accurately determine 9 
non-radial modes, which have RV amplitudes lower than 200 ms$^{-1 }$, a factor of 10 smaller than the observed RV gradient
(see Antoci et al. 2013 for details). The amplitude value of 200  ms$^{-1 }$ also corresponds to the next highest peak in the Fourier spectrum after prewhitening by only the dominant mode and its harmonics. 
Based on our observations, we therefore find it very unlikely that 
non-radial modes are the origin for the large RV differences as a function of line depth. More detailed analyses are 
needed to explore the origin of this phenomenon, however this is 
not the scope of the present paper.



As Mathias et al. (1997) already commented, one has to be careful when using the line bisector technique to extract RVs.
To minimise the scatter when 
averaging the bisector to one RV value, we chose to use only lines showing constant 
amplitude over a large part of the bisector. For the search for solar-like oscillations we have extracted the RV from the following 
lines: Na{\sc i} D$_1$ $\lambda \lambda$ 5895.9\AA, Na{\sc i} D$_2$ $\lambda \lambda$ 
5889.9\AA, Fe{\sc i} Fe{\sc i} blend $\lambda \lambda$ 5139.2/5139.4\AA, Fe{\sc i} Fe{\sc 
i} $\lambda \lambda$ 5107.4/5107.6\AA.  As mentioned above we use lines which were blended by the same element contributing equally to the absorption line. 
All the measurements from a given bisector in the parts where the amplitude was constant 
as a function of depth were averaged. To be consistent with the AAT and McDonald data, 
also in this case the photometrically derived frequencies were prewhitened first. Only 
then the residuals were averaged to a final RV data set, which has a noise level of only 1.2 
ms$^{-1}$ in the Fourier spectrum.

The ESPaDOnS spectra cover the entire visible wavelength region between 3700 and 10480\,\AA. Dravins et al. (1977)
reported a transient emission feature in the Ca{\sc ii} K line. After inspection of all 503 spectra by eye, we do not detect any 
obvious feature which might suggest emission at any given time of the pulsation cycle.

\subsection[]{Interferometry}

\begin{figure}
\begin{center}
\includegraphics[width=8cm]{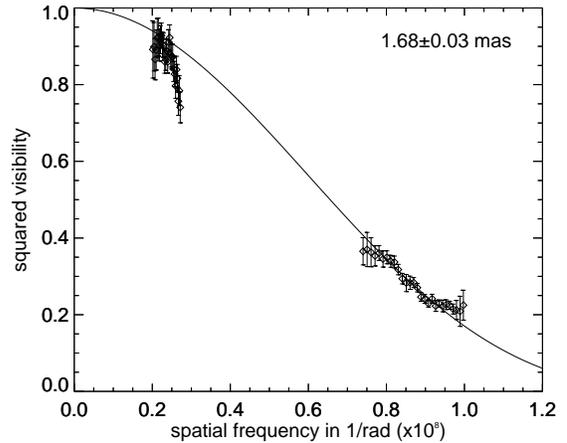}
\caption[]{Squared visibility ($V^2$) versus spatial frequency (projected baseline per unit
wavelength) from the PAVO beam combiner at SUSI. The points show the measured $V^2$ of
$\rho$ Puppis while the error bars show the standard error of mean. The solid line is the
best-fitting limb-darkened (LD) disk model which yields an angular diameter of
1.68 $\pm$ 0.03 mas. }

\label{inter}

\end{center}
\end{figure}

$\rho$ Puppis was also observed interferometrically on May 12, 2010 (8:40-9:22UT) with the 15-m (N3-S1) and the 55-m
(N4-S1) baseline of the Sydney University Stellar Interferometer (SUSI) (Davis et al.
1999) using the PAVO beam combiner (Ireland et al. 2013). The stellar fringe visibility data ($V^2$), shown in Fig.~\ref{inter}, obtained were first calibrated with a
calibrator star, $\zeta$~Puppis, which has a limb-darkened (LD) angular diameter of
0.42 $\pm$ 0.03 mas (Richichi et al. 2005) and were then fitted to a LD disk model with a
SUSI internal calibration tool and a model fitting tool from JMMC, LITPro (Tallon-Bosc
et al. 2008). The linear limb-darkening coefficient (LDC) for the disk model was
obtained from Claret \& Bloemen (2011) with the following parameters; $T_{\rm eff}= $ 7000~K,
$\log g =3.9$ and $[Fe/H]=0.35$. The limb-darkened angular diameter of $\rho$ Puppis obtained
from the model fit is $1.68\pm0.03$ mas, which in combination with the star's HIPPARCOS parallax of $51.33 \pm 0.15$ mas (van Leeuwen 2007) translates into a radius of 3.52 $\pm$ 0.07~R$_{\odot}$.

\section{Analyses}

Frequency analyses were done using the well tested package Period04 (Lens \& Breger 2005). This software uses Fourier and least-square fitting algorithms.

First we performed separate frequency analyses of each individual RV subsets (for AAT and McDonald: chunks; ESPaDOnS: spectral lines), meaning that we prewhitened by the signals at frequencies lower than 20 cd$^{-1}$, known from photometric measurements. These frequencies are: 
$f_1=$ 7.10~cd$^{-1}$, 2$f_1$, 3$f_1$, 4$f_1$, $f_2=$ 6.43~cd$^{-1}$, $f_3=$ 8.82~cd$^{-1}$, $f_4 =$ 8.09~cd$^{-1}$, $f_5=$ 6.52~cd$^{-1}$, $f_6=$ 7.07~cd$^{-1}$, $f_7=$ 6.80~cd$^{-1}$, 2$f_2$, 2$f_3$, $f_1 + f_2$, $f_1 + f_3$, $f_3 - f_1$, $f_3 - f_2$. Detailed analyses of our photometric data are subject to a different paper  (Antoci et al. 2013,  in preparation). In the case of the AAT data we also prewhitened by a signal at 2.00 c/d, which which is certainly not of stellar origin. 

\begin{figure}
\begin{center}
\includegraphics[width=8cm]{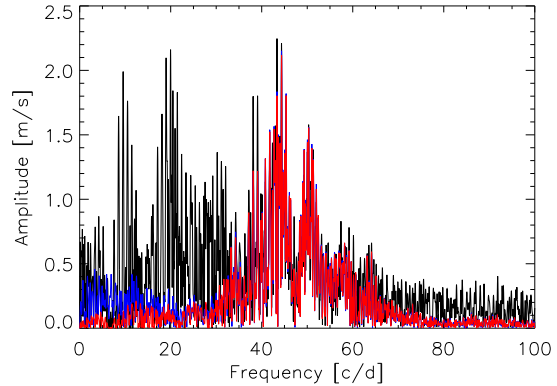}
\caption[]{Fourier spectrum of artificial data with the same sampling (i.e. HJD values) as the AAT data. We simulated stochastic and coherent signals to mimic the solar-like and $\delta$ Sct pulsations, respectively. In black we illustrate the Fourier spectrum after prewhitening by around 40 coherent frequencies (also wrong ones). The blue line represents the Fourier spectrum of the same artificial data set but this time prewhitening by the correct coherent frequencies. In red we show the stochastic signal we used for these simulations. Comparing all there spectra it can be seen that there are only minor changes in amplitude allowing the conclusion that stochastic oscillations would survive the prwhitening process if present.  }

\label{art}

\end{center}
\end{figure}

We then averaged the residuals originating from the different chunks or spectral lines, to a single RV, which was then used for further analyses. We have repeated this procedure for each data set. After combining the data sets we further prewhitened by the remaining signals with frequencies lower than 20 cd$^{-1}$. The results of each data set can be seen in Fig. \ref{resw}, whereas the combined data are displayed in Fig. \ref{modhjd}. From 
tests on artificial data sets with the same time sampling as our observations, containing coherent and non-coherent signals, 
representing the $\kappa$-mechanism driven 
and the stochastic oscillations respectively, we made sure that prewhitening a coherent 
signal does not change a possible non-coherent one. In other words, the lower frequency 
signals observed in $\rho$ Puppis could be removed without affecting the detectability of 
possible solar-like oscillations. Due to the short time-base of the individual data sets most of the 
frequencies derived photometrically cannot be resolved in this case. However, this does not
affect our analysis because we are not interested in the $\delta$ Sct pulsations here, but rather 
want to filter those out to lower the noise and to unveil possible low-amplitude solar-like 
oscillations. Consequently we also removed peaks at low frequencies which were believed of instrumental origin. Figure \ref{art} shows the results of the above mentioned test, depicting in black the Fourier spectrum of the artificial data after prewhitening by the coherent frequencies. To test to what extent this process might change the results we, on the one hand, subtracted `correct' but also `wrong' frequencies. The same figure shows the Fourier spectrum of the same data set but when prewhitening by the `correct' peaks, i.e. the ones used to generate the data set,  in blue and in red the original signal mimicking the solar-like oscillations. It can be seen that the changes are minimal if at all visible, which leads us to the conclusion that our approach of filtering the $\delta$ Sct pulsations does not affect a possible stochastic signal significantly (see figure caption for more details). Further tests can also be found in Antoci et al. (2011, Supplementary Information).   \par

The amplitudes of solar-like oscillations for a star with the parameters of $\rho$ Puppis 
were predicted to be between 1 and 2 ms$^{-1}$ (i.e., $2500-5000$ times lower than the 
amplitude of the dominant $\delta$ Sct pulsation mode!) depending on the approach used 
(Houdek \& Gough 2002 or Samadi et al. 2005). If we use the significance criterion for 
the detection of solar-like oscillations by Kjeldsen \& Bedding (1995) of $S/N = 2.5$, a 
noise level of 0.4 ms$^{-1}$ is needed to detect a signal with a height of 1 ms$^{-1}$. Note that the amplitude of a stochastic signal is not equal to the height of the peak in the Fourier spectrum, as it is the case for a coherent signal. 
If we adopt the significance criterion of Carrier et al. (2007) of $S/N = 3.05$, the noise 
level needs to be at 0.3 ms$^{-1}$. According to the authors, $S/N = 3.05$ is required to 
unambiguously detect solar-like oscillations, with a confidence level of 99\%. \par
However, we can also use the procedure adopted by Kjeldsen et al. (2008) to compute the 
amplitude per radial mode for solar-like oscillations, where (1) the power density spectrum (PDS) is 
convolved with a Gaussian with a FWHM\footnote{Full Width at Half Maximum} equal to four times $\Delta\nu$, the latter being the large frequency separation. (2) We then subtracted the background noise and (3) then calculated the amplitude per oscillation mode by taking the square root of the convolved PDS multiplied by $\Delta\nu/c$, where c is a factor which depends on the spatial response functions, i.e. the visibility of each mode and was set to 4.09, see Kjeldsen et al. (2008) for detailed explanation. Using this procedure we obtained an amplitude per radial mode of 0.5 ms$^{-1}$ . 

Usually when analyzing RV measurements of solar-type stars (e.g. Arentoft et al. (2008), an error per measurement can 
be computed after combining all the RVs extracted from the different spectral chunks. 
These errors can then be used to weigh the data such that less precise measurements 
affect the results less. In the present case, however, better precision has been achieved 
when first prewhitening by the dominant signal at low frequencies and only then combining 
the residuals of the individual RV, as already mentioned. This means that we do not have 
an error per measurement to weigh the data accordingly.  This is because due to the amplitude variability from chunk to chunk the standard deviation, which is usually used, would not reflect the quality of the data but only the deviation from its mean value.

\begin{figure}

\begin{center}

\includegraphics[width=8cm]{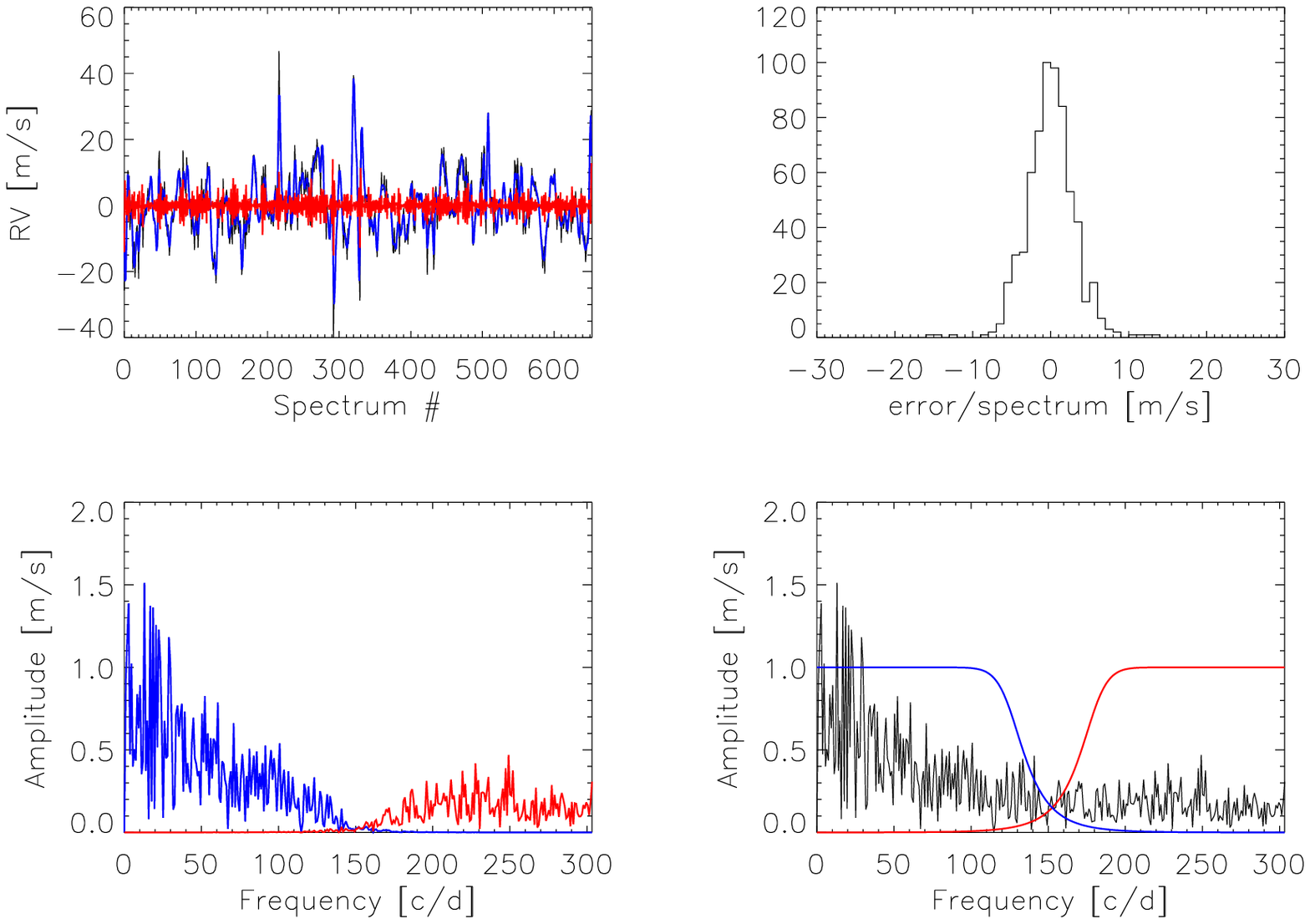}


\includegraphics[width=8cm]{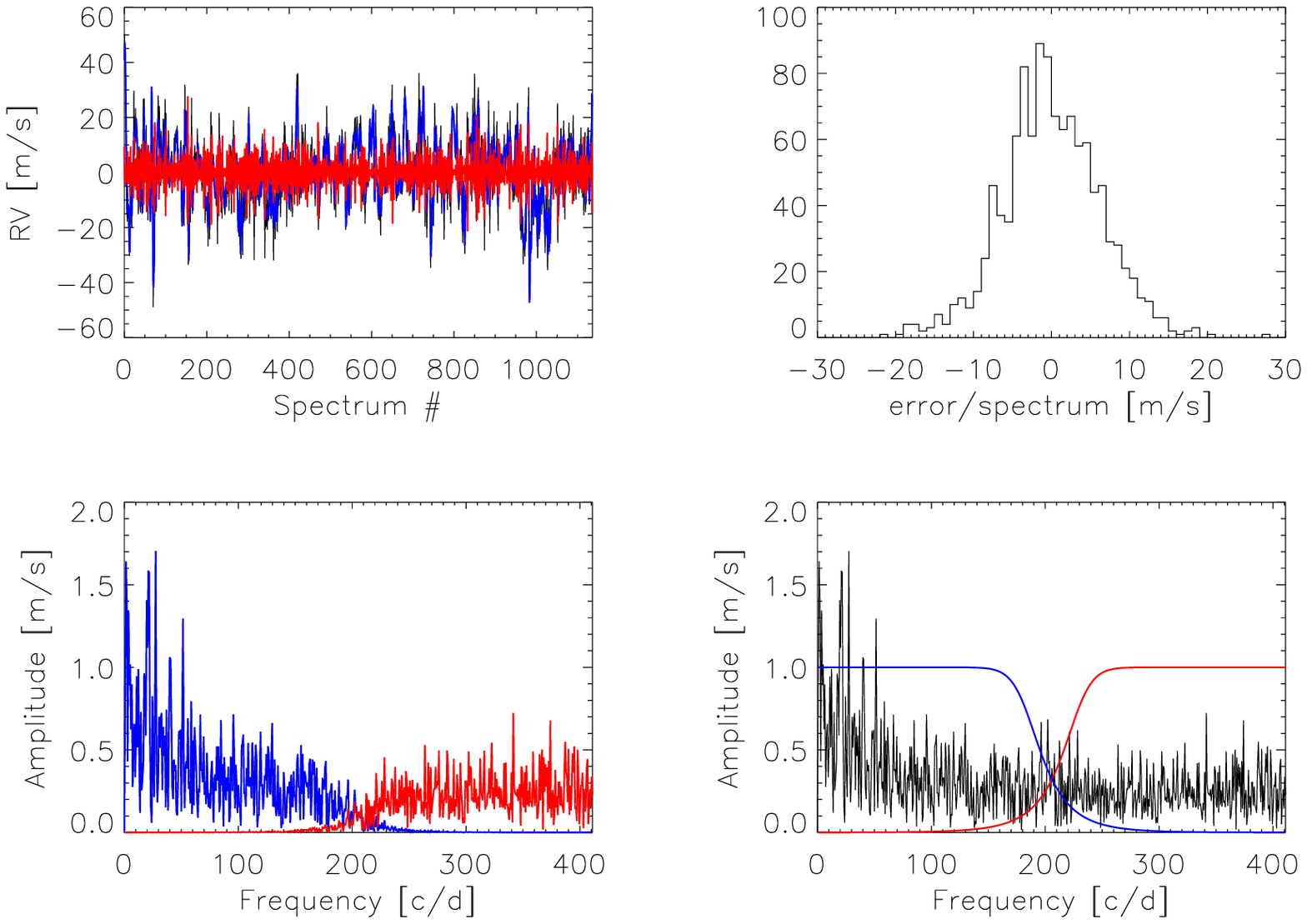}

\caption[]{ High pass filtering using a Butterworth filter to determine the error per measurement for the CORALIE (upper four panels) and AAT data (lower four panels). Upper left panel: observed residuals in black, in red the signal after high pass filtering, in blue the low pass filtered data. Lower left panel: Fourier spectrum of the high (red) and low (blue) pass filtered data. Lower right panel: Fourier spectrum of the observed residuals (black) together with both high and low pass filters. Upper right panel: histogram of the derived uncertainties.  }

\label{butterworth1}

\end{center}

\end{figure}

\begin{figure}

\begin{center}

\includegraphics[width=8cm]{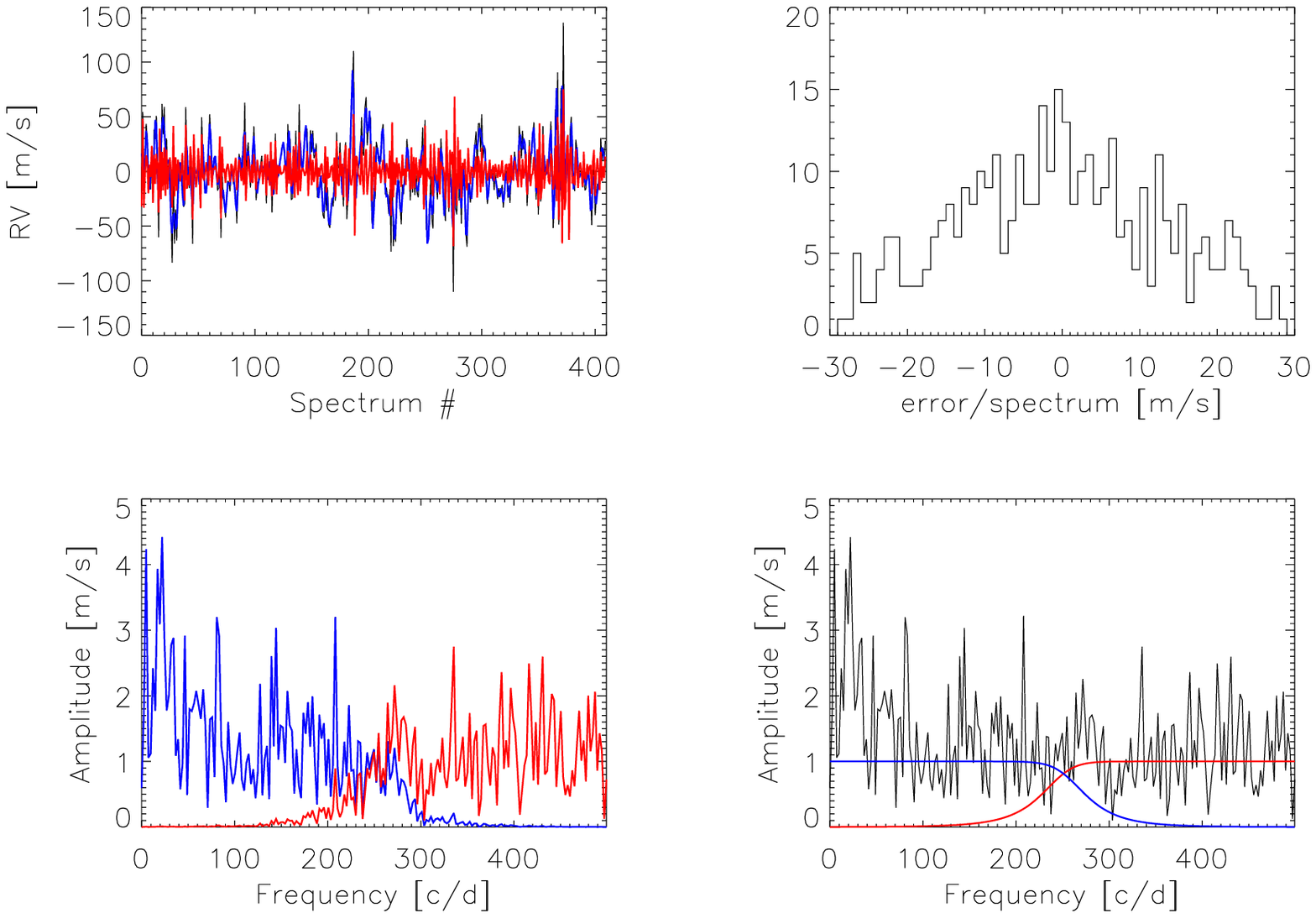}

\includegraphics[width=8cm]{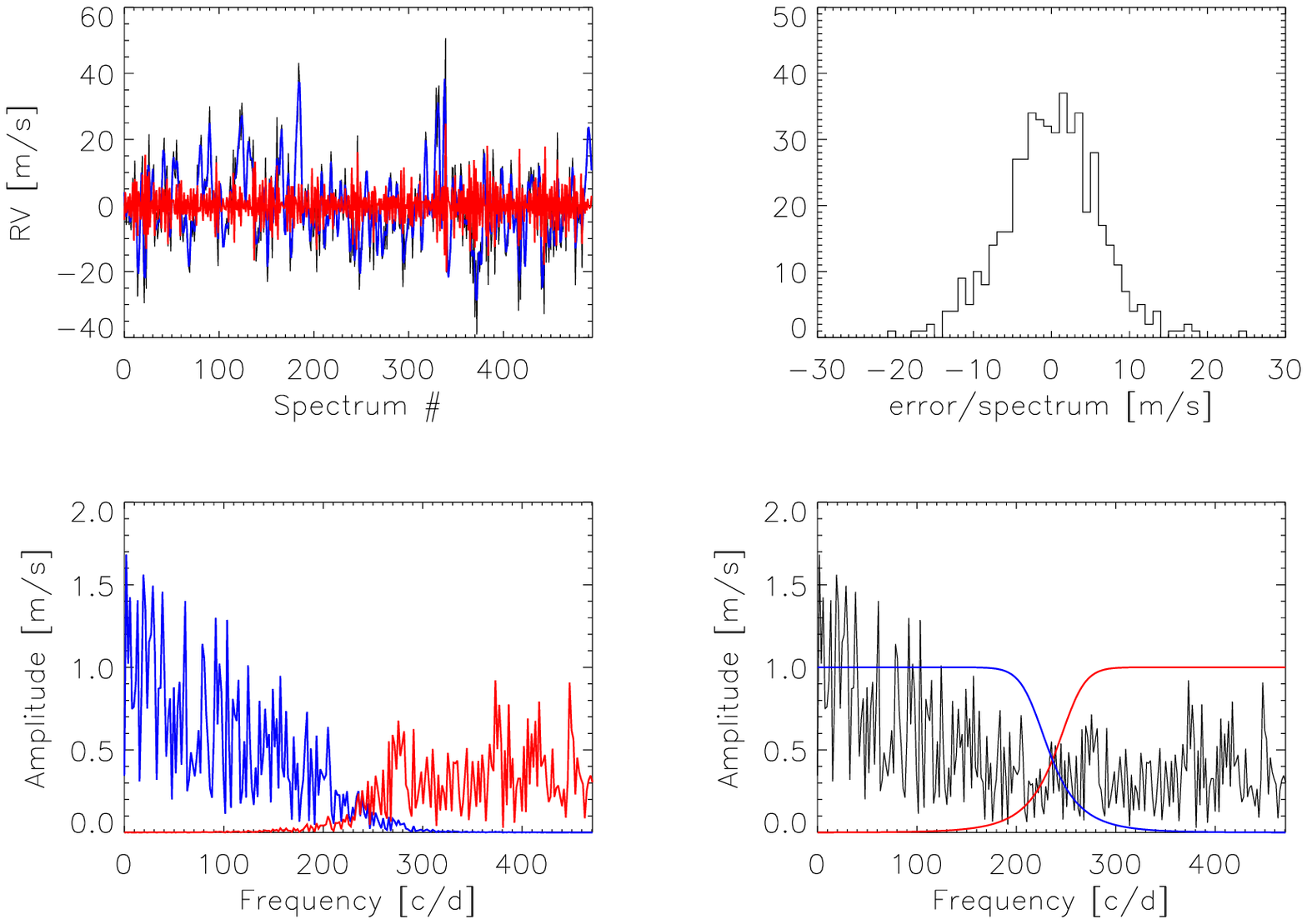}

\caption[]{ High pass filtering using a Butterworth filter to determine the error per measurement for the McDonald data (upper four panels) and ESPaDOnS data (lower four panels). Upper left panel: observed residuals in black, in red the signal after high pass filtering, in blue the low pass filtered data. Lower left panel: Fourier spectrum of the high (red) and low (blue) pass filtered data. Lower right panel: Fourier spectrum of the observed residuals (black) together with both high and low pass filters. Upper right panel: histogram of the derived uncertainties.  }

\label{butterworth3}

\end{center}

\end{figure}

In their extensive observing campaign for Procyon, Arentoft et al. (2008) used 
noise-optimized weights, where they combined the usual weighting scheme ($w_i = 
1/\sigma_i^2$) with the one of Butler et al. (2004)\footnote{a method to weigh down 
outliers by using the condition $\sigma^2_{amp}\sum\limits_{i-1}^{N}\sigma_i^2=\pi$, 
where $\sigma_{amp}$ is the noise level in the amplitude spectrum, N the number of 
measurements, $\sigma_i$ the uncertainty of the velocity measurement.}, but also 
introduced a new method to identify the outliers better and to down-weigh these. To do so 
they employed a high-pass filter to identify which measurements contribute most to the 
signal at very high frequencies. This is possible because, as was shown by e.g. Butler et 
al. (2004), it can be assumed that the amplitude spectrum at high frequencies is dominated by white noise.
Inspired by these methods, we have used the high-pass filtering technique to 
estimate the uncertainties of each measurement in the residual RV curve. To do so we 
used the Butterworth filter deployed in Fourier space, which is available as an IDL 
routine and is described by the following formula:

\begin{equation}
G(\Omega)=\frac{1}{\sqrt{1+(\frac{\Omega}{\Omega_{c}})^{2N}}}
\end{equation}
where $\Omega$ is the frequency in Fourier space, $\Omega_{c}$ the cut-off frequency and 
N the order of the exponent determining the shape of the filter. The gain of this filter was set to unity, meaning that the amplitude of the signal after applying the filter is unchanged (Figs.~\ref{butterworth1} and \ref{butterworth3}). Experimenting with 
different values for the cut-off frequency, we found that the signal higher than 75 
cd$^{-1}$ represents the noise level best. In Figs. \ref{butterworth1} and 
\ref{butterworth3}, lower left-hand panel the Fourier spectra of the high as well as the 
low-pass filter are shown. The latter was not used, but is plotted for completeness. The high-pass  
filter was designed such that the spectral leakage from the 
`unwanted' frequency region is as small as possible. Ideally a passband filter should 
have a so-called {\it brick wall} shape, which should not touch the signal in the pass 
band but filter out the signal in the stop band. The higher the order of the Butterworth 
filter the more a {\it brick wall} shape is approximated. In the present case an order of 
10 was used.  \par

\begin{figure}

\begin{center}

\includegraphics[width=8cm]{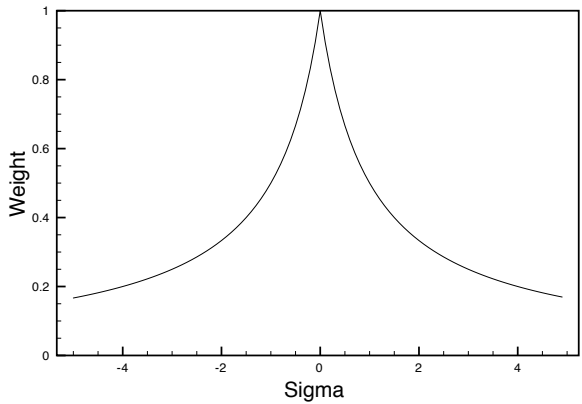}

\caption[]{Function to weight the uncertainties derived as described in the text.}

\label{weight}

\end{center}

\end{figure}

The high-pass filtering was done for each data set separately, because the Nyquist 
frequency for each observing run is different. In Figs.\ \ref{butterworth1} and 
\ref{butterworth3}, we illustrate the filters used as well as the signal after high-pass 
but also after low-pass filtering. The filter was constructed in frequency space, then 
the signal passing through the filter was transformed back to the time domain. This allows  
to determine the amplitude, i.e. the contribution to the signal at high frequencies, which 
mirrors the noise behaviour of the data set, as already mentioned. The same figures also show 
the uncertainties derived from this method as a histogram, giving a clear idea which are 
the better data sets. For further analyses we then computed a mean error and weighted the 
measurements using the function displayed in Fig. \ref{weight}.

\begin{figure}

\begin{center}

\includegraphics[angle=270, width=9cm]{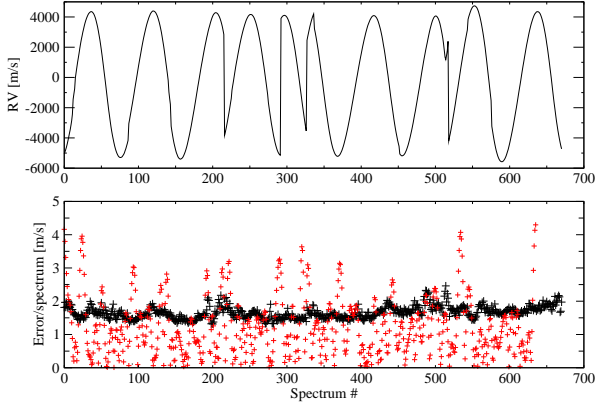}

\caption[]{Upper panel: RV curve observed with CORALIE. Lower panel: Errors per spectrum delivered  by the pipeline when extracting the RV (black) and the errors determined from the Butterworth filter in red. Note that the x-axes is not time but number of measurement. } 

\label{coraliedata}

\end{center}

\end{figure}

\begin{figure}

\begin{center}

\includegraphics[width=14cm]{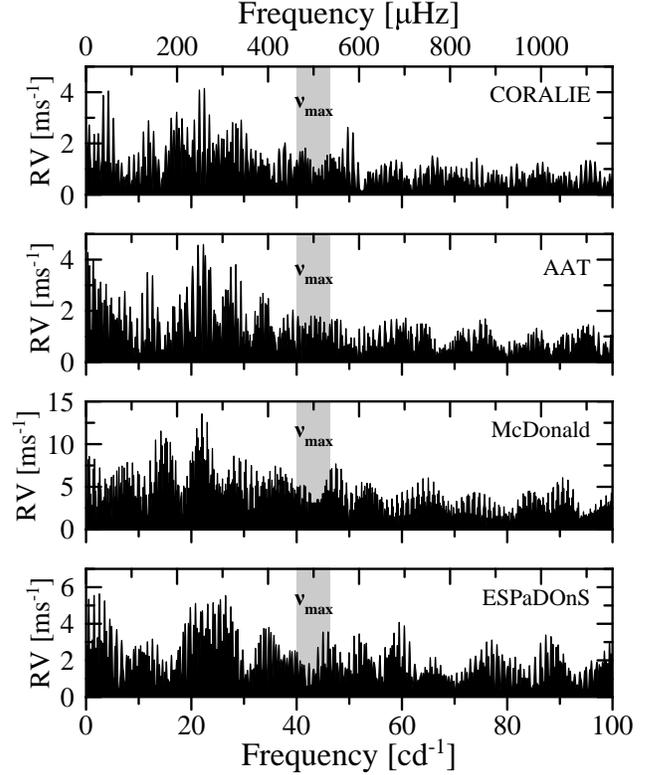}

\caption[]{Fourier spectra for all individual data sets, using 
noise-optimized weights. The grey region indicates where the frequency of maximum power $\nu_{max}\sim 43$~cd$^{-1}$ is expected (see text for details).   }

\label{resw}

\end{center}

\end{figure}

\begin{figure}

\begin{center}

\includegraphics[width=9cm]{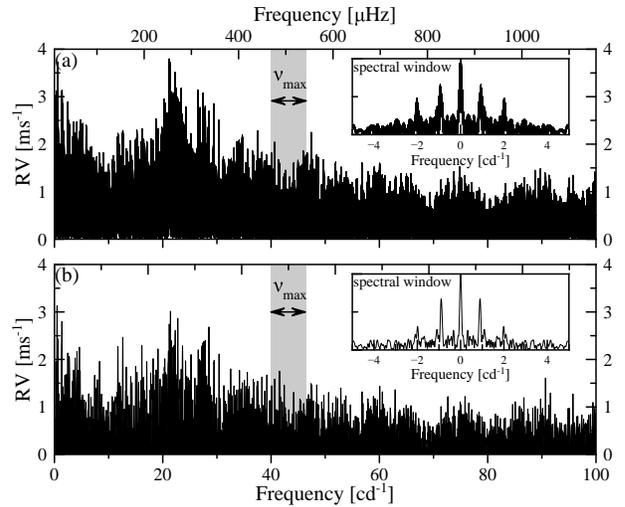}

\caption[]{ Fourier spectrum of all combined data, with the original HJD, hence the name `original' data in panel (a). In grey we depict the region were $\nu_{max}$ is expected. A solar-like mode at a given frequency is damped and re-excited at a different phase, which means preserving the phase information is not absolutely necessary as in the case of a coherent signal. Therefore the gaps can be removed without altering the result of interest. In panel (b) we depict the `stacked' data where gaps longer than 2 days were removed.}

\label{modhjd}

\end{center}

\end{figure}

When inspecting each RV curve prior to the filtering only very obvious outliers were 
omitted. Even if there were more points with lower quality we chose to down-weigh, rather 
than deleting them. In case of the CORALIE data, displayed in Fig.~\ref{coraliedata} 
together with the uncertainties, the errors per measurement were delivered together with 
the RV data, and were determined using the CORALIE pipeline. Sometimes such pipelines can 
underestimate the errors. At least that is what seems to be the case if we compare the 
pipeline errors to the error estimated by using the high-pass-filtering for the individual measurements. 
When using the formula given by Kjeldsen \& Bedding (1995) to estimate the error 
per spectrum by measuring the noise level in the power spectrum\footnote{The noise level was calculated in the frequency region between 150 to 200 cd$^{-1}$ (1735 -- 2314~$\mu Hz$).}:
\begin{equation}
 \sigma_{PS}=4\sigma_{rms}^2/N
\end{equation}
where $\sigma_{PS}$ is the noise level in the power spectrum, N the number of 
measurements and $\sigma_{rms}$ their rms scatter, we obtained for the measured noise $\sigma_{PS}\approx$
0.25~m$^2$/$s^2$ (noise in amplitude $\approx$ 0.5~ms$^{-1}$), hence a $\sigma_{rms}$ of $\sim$ 6.5 ms$^{-1}$. 
This is in better agreement with what we obtained from the high-pass filtering 
(i.e. 3 ms$^{-1}$) than with the values delivered from the pipeline 
(Fig.~\ref{coraliedata}). Also in this case a slight dependence of the errors on the pulsation cycle can be detected, even if more subtle than in the AAT data. For further analyses, we therefore used the errors derived from 
the Butterworth filter illustrated in Fig.~\ref{butterworth1}. The Fourier spectrum of 
the residuals is displayed in Fig.~\ref{resw}. \par

For the AAT data, 104 residual RV curves were combined by averaging the measurements 
 to one final RV curve. This resulted in a noise level of 0.9 ms$^{-1}$ in 
the Fourier spectrum when using the noise-optimized point weights as described above (see 
Fig. \ref{butterworth1} and Fig. \ref{resw}). The same was done for the McDonald data 
set, by combining the RV curves of 27 different chunks yielding a noise level of 2.4 
ms$^{-1}$. The ESPaDOnS RVs were derived by using the line bisectors as described above. 
Also in this case the dominant signals at low frequencies were first removed and only 
then the residuals combined. The noise level in the Fourier spectrum (see 
Fig.~\ref{resw}) is 1.2 ms$^{-1}$. \par


Finally all the data sets were combined, lowering the noise level down to 0.4 ms$^{-1}$ 
when using the weighted data (upper panel of Fig. \ref{modhjd}). For comparison, the 
unweighted data result in a somewhat higher noise of 0.5 ms$^{-1}$. Even if the 
difference may not be huge, in terms of significance criterion it helps to set a lower 
limit. In all panels of the same figure, the grey area indicates where the 
frequency of maximum power is expected. To determine $\nu_{max}$ we can on the one hand use $Q = P \sqrt{\frac{\rho}{\rho_{\odot} } }$, and on the other $\Delta \nu=\Delta\nu_{\odot}\sqrt{\frac{\rho}{\rho_{\odot}}}$, where $Q$ is the pulsational constant (0.033 days for the radial fundamental mode in $\delta$ Sct stars, e.g. Breger 2000) and P the period of the radial fundamental mode (0.14088 days for $\rho$ Puppis, see Antoci et al. (2013) for details). Using the solar value  $\Delta\nu_{\odot}$ of 11.67 cd$^{-1}$ (135.1 $\mu Hz$, Huber et al. 2011) we derive an expected large frequency separation of 2.73 cd$^{-1}$ (31.6 $\mu Hz$). Further, we can apply the empirical scaling relation $\nu_{max}-\Delta\nu$ in the form: 
$\Delta\nu=\alpha(\nu_{max}/\mu Hz)^\beta$, where $\alpha$ varies between 0.20 and 0.25 and $\beta$ between 0.779 and 0.811 depending on the literature (Huber et al. 2011 and references therein). Based on the large frequency separation mentioned above and with $\alpha=0.263$ and $\beta=0.772$ (Stello et al. 2009) we estimate the frequency of maximum power $\nu_{max}$ to be $\sim 43\pm2$~cd$^{-1}$ ($\sim 498\pm23~\mu$Hz). Note that using other values for $\alpha$ and $\beta$ as described in Huber et al. (2011) will  shift the predicted $\nu_{max}$ to slightly higher frequencies.

Solar-like oscillations are damped and re-excited at the same frequency but with 
different phases. This allows us to collapse the HJD of our data, such that the sampling 
of each data set is preserved but the gaps longer than 2 days removed (see, e.g., Kallinger \& Matthews 2010). This procedure, of course, changes the window function, suppressing the annual aliases (see Fig. \ref{modhjd}). The trade-off is lower frequency resolution, which in the present case is less 
important. The results of the `stacked' data are illustrated and compared with the 
`original' data in Fig. \ref{modhjd}. In both cases there is no obvious signal that could 
be interpreted as solar-like oscillations, where $\nu_{max}$ is expected, indicated by the grey area. This may lead to the conclusion that we are unable to detect solar-like oscillations in  $\rho$ Puppis within the limitations of our data set. However, 
there is clearly a signal at lower frequencies around 22 and 29 cd$^{-1}$, which can, if 
intrinsic to the star, either be due to (a) imperfect prewhitening of the 3rd and 4th harmonic of the dominant mode (b) combination frequencies from the $\delta$ 
Scuti pulsation or (c) independent high order $\kappa$ mechanism driven pulsation. Note that the signals are statistically not significant, considering the significance criterion of S/N=4 used for $\kappa$ driven pulsators, i.e. for coherent oscillators (Breger et al. 1999). \par

Nevertheless let us assume as an example the possibility that we do observe solar-like 
oscillations in $\rho$ Puppis, not with a maximum amplitude at the predicted $\nu_{max}$ 
of   $\sim$ 43~cd$^{-1}$ ($\sim498~\mu$Hz), but at around $\sim 29 \pm2$~cd$^{-1}$ ($\sim335\pm24~\mu$Hz) where power is observed. If we apply 
the relation from Kjeldsen \& Bedding (1995):
\begin{equation}
\begin{aligned}
 \frac{\nu_{max}}{\nu_{max\odot}}  & =    \frac{M/M_{\odot}}{(R/R_{\odot})^2\sqrt{(T_{\rm eff}/5777K)}} 
\end{aligned}
\end{equation}
and use the radius derived from interferometry (R = $3.52 \pm0.07$~R$_{\odot}$) together with the 
T$_{\rm eff}$ of $6900\pm150$~K we obtain a mass of $\sim 1.5\pm0.1$~M$_{\odot}$. For completeness we did the same exercise assuming that the signal at 
$\sim$ 22 cd$^{-1}$ ($\sim254~\mu Hz$) is $\nu_{max}$. To fit this value for the same radius and temperature 
as above, a mass of $\sim1.1\pm0.1$~M$_{\odot}$ is required, which is definitely inconsistent with the mass of a Pop. I $\delta$ Scuti star. We find it most likely that the power excess, even if not statistically significant, originates from the maybe imperfect prewhitening process, as the values are around the 3rd and 4th harmonic of the dominant mode.

To conclude, we have no evidence of solar-like oscillations in $\rho$ Puppis at the 
predicted frequency of maximum power of $43\pm2$~cd$^{-1}$  with a peak-amplitude per radial mode higher than 0.4~ms$^{-1}$, following the procedure of Kjeldsen et al. (2008).

\section{Conclusions}

The theoretical prediction for the presence of solar-like oscillations of $\delta$ Sct stars motivated us to organise a high-resolution high-precision spectroscopic multi-site campaign involving four different observatories for the $\delta$ Sct star $\rho$ Puppis. Two of the data sets were acquired using the iodine-cell technique, while the other two made use of ThAr lamps for wavelength calibration. Our efforts resulted into a noise level in the Fourier spectrum of only 0.4 ms$^{-1}$ of the residual RV data, after prewhitening by the frequencies lower than 20~cd$^{-1}$ and weighing probable outliers, which makes it the best time-resolved  spectroscopic data set ever obtained for a $\delta$ Sct star. We conclude that the iodine cell technique can also be used for $\delta$ Sct stars to derive accurate RVs, even with a high amplitude of pulsation, especially when using the sliding-window technique.  When combining the individual RV measurements for each spectrum, however, we could not just combine these as it is usually done for planet hunting and solar-like oscillators, but had to first subtract the high amplitude signal. That was necessary because of strong amplitude variations from line to line. To estimate the error per measurement we used high-pass filtering.  \par

The $\delta$ Sct star $\rho$ Puppis is the best target for the search of solar-like oscillations from ground: it is bright (V=2.81), the dominant mode is known to be  the radial fundamental mode, it is metal-rich and has therefore many lines containing RV information, and for a higher metallicity the amplitude of solar-like oscillations are expected to be higher (Houdek \& Gough 2002, Samadi et al. 2005).
Our data enabled us to confirm the findings of Yang et al. (1987) and Mathias et al. (1997), who reported different amplitudes in lines of different elements up to 2000  ms$^{-1}$. We also find differences in the RV amplitudes as a function of line depth. We have constructed bisectors which clearly show velocity gradients as large as 2000 ms$^{-1}$. These can be excluded to originate from non-radial modes, as the amplitude of the strongest non-radial mode is expected to be below 200 ms$^{-1}$, i.e., less than one-tenth of the measured effect. A likely scenario is that we observe the running wave character of the dominant mode as described in Mathias et al. (1997, and references therein). However to confirm this interpretation more detailed analyses are needed which are not the scope of this present paper. We also do not find any (transient) emission feature in the Ca {\sc ii} K line, as suggested by Dravins et al. (1977), consistent with Mathias et al. (1999). 

A new interferometric measurement of $\rho$ Puppis yielded an angular diameter of 1.68 $\pm$ 0.03~mas, which translates into a radius of 3.52 $\pm 0.07$~R$_{\odot}$.\par
The dominant mode of pulsation is the radial fundamental mode, which allows us to directly determine the expected large frequency separation $\Delta \nu$. Using the $\Delta \nu$-$\nu_{max}$ scaling relation we predict the solar-like oscillation to have a maximum amplitude at $\sim43 \pm 2$~cd$^{-1}$  ($\sim498\pm23~\mu Hz$). Given the noise level of 0.4 ms$^{-1}$ we can exclude the presence of significant peaks higher than 1.2~ms$^{-1}$
using the significance criterion of S/N=3.05 introduced by Carrier et al. (1997). Finally, following Kjeldsen et al. (2008), we exclude the presence of solar like oscillations with a peak-amplitude per radial mode of 0.5 ms$^{-1}$.\par
Recently, Antoci et al. (2011) reported the first evidence for the presence of solar-like oscillations in the $\delta$ Sct Am star HD 187547. Their interpretation was supported by the striking similarities between the observed signal and the properties of solar-like oscillations. New observations of this star, however make the picture somehow more complicated, as additional data from the {\it Kepler} spacecraft indicate that the mode lifetimes of the modes interpreted as being stochastically excited are longer than duration of the data sets. This is not consistent with stochastically  excited modes as observed in sun-like stars. The fact that $\rho$ Puppis and hundreds other {\it Kepler} $\delta$ Sct stars do not show solar-like oscillations may lead to the conclusion that the theory predicting this type of pulsation is still incomplete, especially because all solar-like stars observed long enough show stochastic oscillations. The question, however, is if this is also to be expected for the $\delta$ Sct stars, mainly because these stars are certainly not as homogeneous as solar-type stars, e.g. when it comes to the distribution of {rotational velocities, e.g. 44 Tau, Zima et al. (2009) and Altair, e.g. Monnier et al. (2007)}. In the case of HD 187547, there is a smooth transition between the $\delta$ Sct modes and the modes interpreted as solar-like oscillations, as predicted from the $\Delta \nu$-$\nu_{max}$ scaling relations, suggesting similar time scales and therefore an interaction that might also explain the observed mode lifetimes. This is not predicted to be observed in $\rho$ Puppis. So far there is no theoretical work to investigate to which extent and how the $\kappa$ mechanism may influence the acoustic oscillations driven by the turbulent motion of convection, but the fact that even in the Sun the $\kappa$ mechanism plays a role (Balmforth 1992), even if minor, shows the necessity to consider this. Maybe solar-like oscillations are only excited in $\delta$ Sct stars where the two regions overlap, because the convection might not be effective enough to excite these pulsations alone but gets additional power from the $\kappa$ mechanism which acts at similar time scales. At the moment this hypothesis is purely speculative but needs to be tested, both from theoretical point of view but also observationally. 

For results of our photometric campaign and modelling of the $\delta$~Scuti pulsations of $\rho$ Puppis we refer to Antoci et al. (2013).

\section*{Acknowledgments}

V.A. and G.H. were supported by the Austrian Fonds zur F\"orderung der wissenschaftlichen 
Forschung (Projects P18339-N08 and P20526-N16). Funding for the Stellar Astrophysics 
Centre (SAC) is provided by The Danish National Research Foundation. The research is 
supported by the ASTERISK project (ASTERoseismic Investigations with SONG and {\it Kepler}) 
funded by the European Research Council (Grant agreement no.: 267864). GH acknowledges support by the Polish NCN grant 2011/01/B/ST9/05448. We thank 
Michel Endl for helping with organising some observations, G\"unter Houdek 
for valuable discussions as well as Nadine Manset, Kristin Fiegert and Stephen Marsden for assistance with some of the observations.

\label{lastpage}

\end{document}